\documentclass[aps,twocolumn,showpacs,superscriptaddress,amsmath,amssymb,amsfonts,floatfix,longbibliography]{revtex4-1}

\usepackage[T1]{fontenc} 
\usepackage{graphicx}
\usepackage{float}
\usepackage{dcolumn}
\usepackage{bm}
\usepackage{amssymb}
\usepackage{microtype}
\usepackage{xfrac}
\usepackage{gensymb}
\usepackage[most]{tcolorbox}
\usepackage{xcolor}
\usepackage{multirow}
\usepackage{enumitem}
\usepackage{natbib,hyperref}
\usepackage{graphicx}
\usepackage{ulem}

\hypersetup{
	colorlinks=true, 
	linkcolor=blue,  
}
\newcommand{\PreserveBackslash}[1]{\let\temp=\\#1\let\\=\temp}

\newcolumntype{C}[1]{>{\PreserveBackslash\centering}p{#1}}

\begin{document}

\title{Entropy engineering and tunable magnetic order in the spinel high entropy oxide}

\author{Graham H.J. Johnstone}

\affiliation{Department of Physics \& Astronomy, University of British Columbia, Vancouver, BC V6T 1Z1, Canada}
\affiliation{Stewart Blusson Quantum Matter Institute, University of British Columbia, Vancouver, BC V6T 1Z4, Canada}

\author{Mario U. Gonz\'alez-Rivas}

\affiliation{Department of Physics \& Astronomy, University of British Columbia, Vancouver, BC V6T 1Z1, Canada}
\affiliation{Stewart Blusson Quantum Matter Institute, University of British Columbia, Vancouver, BC V6T 1Z4, Canada}

\author{Keith M. Taddei}
\affiliation{Neutron Scattering Division, Oak Ridge National Laboratory, Oak Ridge, Tennessee 37831, USA}

\author{Ronny Sutarto}
\affiliation{Canadian Light Source, Saskatoon, Saskatchewan S7N 2V3, Canada}

\author{George A. Sawatzky}
\affiliation{Department of Physics \& Astronomy, University of British Columbia, Vancouver, BC V6T 1Z1, Canada}
\affiliation{Stewart Blusson Quantum Matter Institute, University of British Columbia, Vancouver, BC V6T 1Z4, Canada}

\author{Robert J. Green}
\affiliation{Department of Physics and Engineering Physics, University of Saskatchewan, Saskatoon, SK S7N 5E2, Canada}
\affiliation{Stewart Blusson Quantum Matter Institute, University of British Columbia, Vancouver, BC V6T 1Z4, Canada}

\author{Mohamed Oudah}
\affiliation{Department of Physics \& Astronomy, University of British Columbia, Vancouver, BC V6T 1Z1, Canada}
\affiliation{Stewart Blusson Quantum Matter Institute, University of British Columbia, Vancouver, BC V6T 1Z4, Canada}

\author{Alannah M. Hallas}
\email[Email: ]{alannah.hallas@ubc.ca}
\thanks{G.H.J.J. and M.U.G.-R. contributed equally to this work.}
\affiliation{Department of Physics \& Astronomy, University of British Columbia, Vancouver, BC V6T 1Z1, Canada}
\affiliation{Stewart Blusson Quantum Matter Institute, University of British Columbia, Vancouver, BC V6T 1Z4, Canada}

\begin{abstract}
\begin{center}
{\normalsize\textbf{ABSTRACT}}\\    
\end{center}
Spinel oxides are an ideal setting to explore the interplay between configurational entropy, site selectivity, and magnetism in high entropy oxides. In this work we characterize the magnetic properties of the spinel (Cr,Mn,Fe,Co,Ni)$_3$O$_4$ and study the evolution of its magnetism as a function of non-magnetic gallium substitution. Across the range of compositions studied here, from 0\% to 40\% Ga, magnetic susceptibility and powder neutron diffraction measurements show that ferrimagnetic order is robust in the spinel HEO. However, we also find that the ferrimagnetic order is highly tunable, with the ordering temperature, saturated and sublattice moments, and magnetic hardness all varying significantly as a function of Ga concentration. Through x-ray absorption and magnetic circular dichroism, we are able to correlate this magnetic tunability  with strong site
selectivity between the various cations and the tetrahedral and octahedral sites in the spinel structure. In particular, we find that while Ni and Cr are largely unaffected by the substitution with Ga, the occupancies of Mn, Co, and Fe are each significantly redistributed. Ga substitution also requires an overall reduction in the transition metal valence, and this is entirely accommodated by Mn. Finally, we show that while site selectivity has an overall suppressing effect on the configurational
entropy, over a certain range of compositions, Ga substitution yields a striking increase in the configurational entropy and may confer additional stabilization. Spinel oxides can be tuned seamlessly from the low-entropy to the high-entropy regime, making this an ideal platform for entropy engineering.
\end{abstract}

\maketitle

\section{Introduction}

Since their initial discovery in 2015~\cite{rost2015entropy}, the rapidly growing field of high entropy oxides (HEOs) has attracted significant interest from chemists, physicists, and materials scientists alike. There is no universally agreed upon definition for what constitutes an HEO, but in general terms they are crystalline oxides with a high degree of configurational entropy due to the solid solution of multiple (in many cases five) metal cations sharing a single sublattice~\cite{zhang2019review,oses2020high,musico2020emergent,sarkar2020high,mccormack2021thermodynamics}. In certain cases, this configurational entropy is so large that it dominates the free energy landscape at the high temperatures where materials synthesis occurs, such that the resulting single phase solid solution is actually selected by entropic rather than enthalpic forces. HEOs have been demonstrated to have highly tunable properties and enhanced structural stability which lends credence to the idea that they may one day find widespread use in applications such as reversible energy storage~\cite{berardan2016room,sarkar2018high,sarkar2019high} or catalysis~\cite{sun2021high}. One question that has elicited significant interest is the degree to which the ideal configurational disorder is actually achieved in real HEOs or whether it is significantly reduced due to short-range ordering or clustering, and the extent to which these effects might alter the structural, magnetic, and electronic properties.   

The cubic spinel family of materials, with the general chemical formula $AB_2$O$_4$~\cite{sickafus1999structure,zhao2017spinels}, provide a particularly interesting setting in which to explore the interplay between configurational disorder and structure-function relationships in HEOs. First, the spinel structure offers not one but two cation sublattices: the tetrahedrally coordinated $A$ site that forms a diamond lattice and the octahedrally coordinated $B$ site that forms a pyrochlore lattice. In the extreme limit where both sublattices are occupied by a mixture of cations a dramatic enhancement to configurational entropy can be expected. Spinels can also exist as both normal (valence ordered) or inverted (mixing of valences between $A$ and $B$ site)~\cite{verwey1947physical,o1983simple}, they are known to exhibit oxygen vacancies~\cite{song2012role,torres2014oxygen,zhao2017spinels}, and they are susceptible to short-range cation ordering~\cite{o2017inversion,liu2019unified}. Each of these factors could enhance or diminish the overall degree of configurational entropy. In fact, the notion of entropy stabilization in spinels vastly predates the concept of HEOs in the modern sense. For example, in a series of papers, Navrotsky \textit{et al.} show that even for binary spinel systems such as MgAl$_2$O$_4$, mixing of the cation sites provides the necessary configurational entropy to stabilize the spinel phase in relation to the component oxide phases, which are favored by enthalpy~\cite{navrotsky1967thermodynamics,navrotsky1968thermodynamics,navrotsky1969thermodynamics}.

\begin{figure*}[!ht]
  \includegraphics[width=17cm]{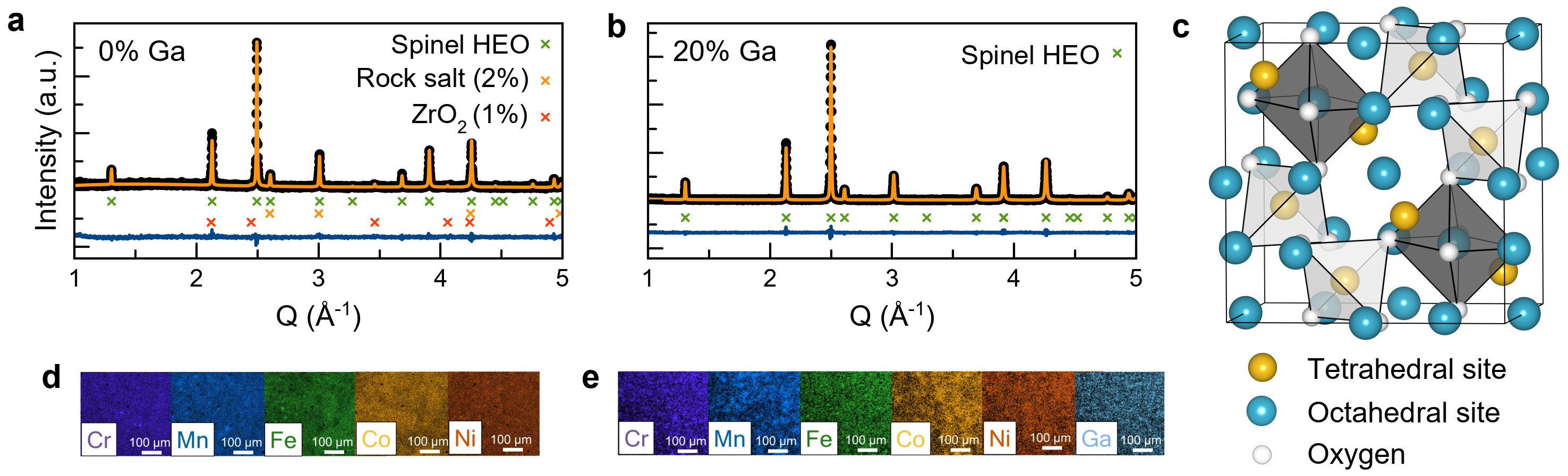}
  \caption{Structural characterizations of the HEO spinels. Rietveld refinement (yellow lines) of the powder x-ray diffraction data (black circles) on the (\textbf{a}) 0\% Ga and (\textbf{b}) 20\% Ga samples confirm the cubic spinel phase  (residual indicated by the blue line). The Bragg peak positions for the $Fd\bar{3}m$ spinel structure are indicated by the green crosses. Minor rock salt and zirconia impurities in the 0\% sample are indicated by yellow and orange crosses, respectively. (\textbf{c}) The cubic spinel structure is made up of tetrahedrally coordinated (yellow) and octahedrally coordinated (blue) cation sites. Energy dispersive x-ray spectroscopy mapping for the (\textbf{d}) 0\% and (\textbf{e}) 20\% Ga samples reveal that the elemental distribution in these spinel HEOs are largely homogeneous on the micron scale, with a higher degree of clustering evident in the Ga-substituted sample.}
  \label{XRD_EDX}
\end{figure*}

The spinel family also offers a rich opportunity to study the intersection of magnetism with high configurational disorder. Previous studies on the prototypical rock salt HEO, (Mg,Co,Ni,Cu,Zn)O, have revealed that long range antiferromagnetic order can occur in high entropy systems, even when a high concentration of the sites (40\%) are occupied by non-magnetic ions \cite{ChemOfMat-Rocksalt-magnetism}. This finding raises natural questions about the versatility and robustness of magnetic order in HEOs more generally. In particular, are there limitations on the types of magnetically ordered states that can be obtained in a high entropy material and how does magnetic dilution interface with the strong configurational disorder? In contrast to the much-studied rock salt HEO, in the spinel structure it becomes possible to synthesize a five-component material (Cr,Mn,Fe,Co,Ni)$_3$O$_4$ composed entirely of magnetic transition metals \cite{s-HEO-first-report}. Previous studies on this material revealed long-range magnetic order \cite{s-HEO-first-report, Spinel-HEO-Cieslak-FiM-Mossbauer,Spinel-HEO-Mao-FiM} and remarkable electrochemical properties, raising interest in their potential for energy applications \cite{Spinel-HEO-Grzesik-diffusion,Spinel-HEO-Huang-Lithiation,Spinel-HEO-Hygar-Seebeck,Spinel-HEO-Nguyen-batteries,Spinel-HEO-Talluri-supercapacitor,Spinel-HEO-Wang-batteries,Spinel-HEO-Wang-water}. Additionally, spinels present the possibility for site selectivity stemming from competing enthalpy and crystal field effects, further enriching the complexity of their magnetic properties~\cite{sarkar2021comprehensive}.

In this paper, we attempt to disentangle the effects of configurational disorder and magnetic dilution on the magnetic ground state of the spinel HEO (Cr,Mn,Fe,Co,Ni)$_3$O$_4$ as a function of non-magnetic Ga substitution. Gallium is the ideal candidate for magnetic dilution in this family, as in addition to carrying no magnetic moment, it is known to form ternary spinel phases with most of the ions present \cite{Casado-MnGa2O4,Ghose-FeGa2O4, Nakatsuka-CoGa2O4-ref,Arean-NiGa2O4}, and, further, its ionic radius is similar to that of the $3d$ transition metals and therefore produces negligible chemical pressure effects. We characterize the effect of this magnetic dilution through an ensemble of measurements with sensitivity to varying length scales (long and short-range order) and degrees of freedom (magnetic and structural). We find that long range ferrimagnetic order is robust across all Ga concentrations. However, the ordering temperature, magnetic hardness, and saturation moment all vary sensitively with composition, resulting in a highly tunable magnetic system. Locally, Ga-substitution enhances the configurational disorder of the spinel HEO system, with the more versatile ions (Mn, Fe, and Co) accommodating for charge imbalances. Spinels are therefore found to be an exemplary system to study the interactions between the configurational disorder, magnetic dilution, and site selectivity.

\begin{figure*}[!ht]
  \centering
  \includegraphics[width=15cm]{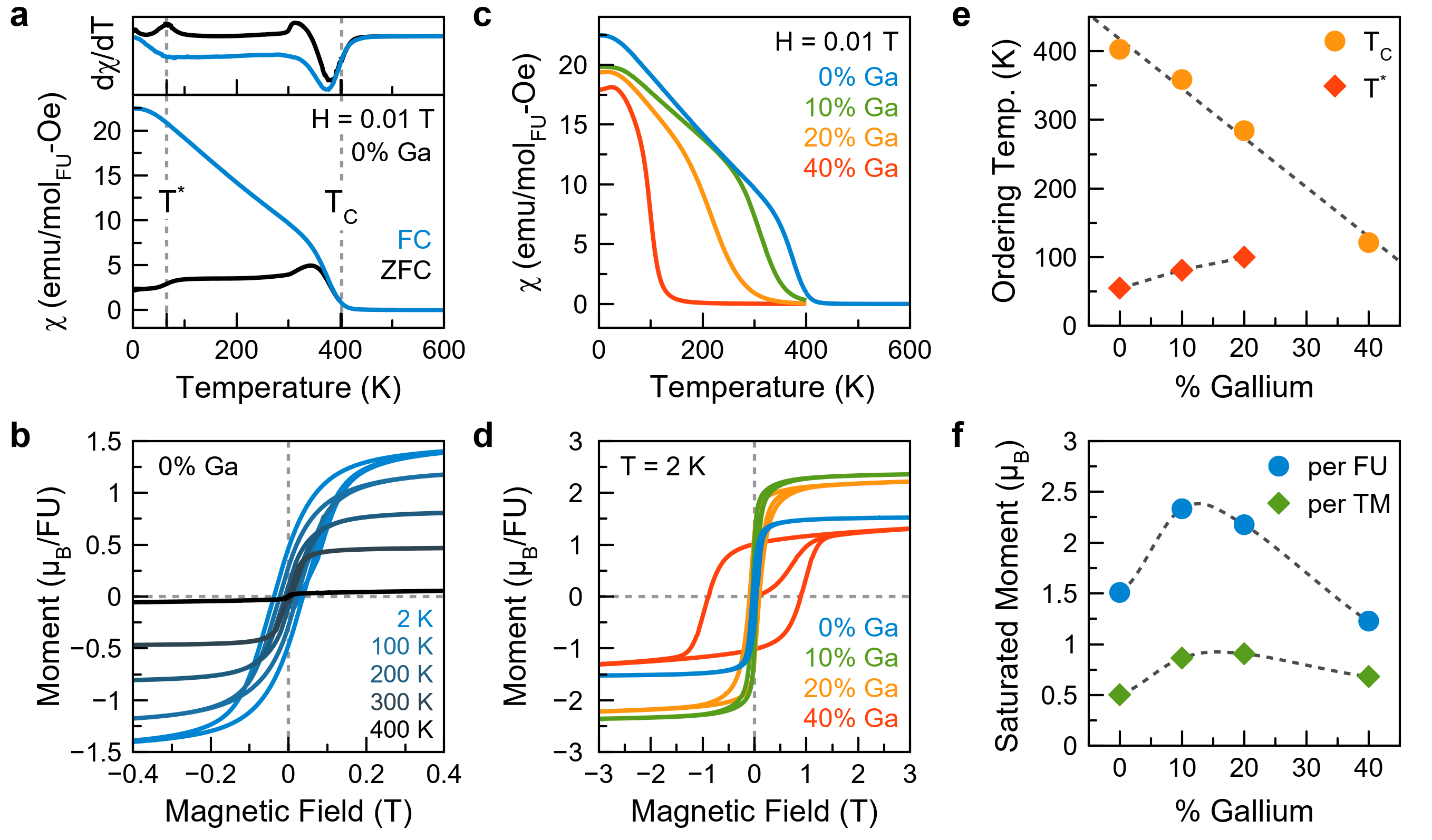}
  \caption{Summary of magnetic property measurements performed on the HEO spinels. (\textbf{a}) Magnetic susceptibility and (\textbf{b}) magnetization of (Cr,Mn,Fe,Co,Ni)$_3$O$_4$ revealing the onset of ferrimagnetic order at $T_C = 403$~K followed by a second anomaly at $T^* = 55$~K and a small hysteresis within the ordered state with a saturated moment of $1.52~\mu_{\text{B}}$ per formula unit (FU). (\textbf{c}) The field-cooled magnetic susceptibility and (\textbf{d}) $T=2$~K magnetization for the Ga-substituted HEO-spinels showing the suppression of magnetic order and a widening hysteresis with increasing Ga-concentration. The variation of (\textbf{e}) $T_C$ and $T^*$ and (\textbf{f}) the saturated moment per formula unit and per magnetic transition metal (TM) with Ga-concentration.}
  \label{Susceptibility_Summary}
\end{figure*}

\section{Results and discussion}

\subsection{Crystal structure}

Polycrystalline samples of the spinel high entropy oxide (Cr,Mn,Fe,Co,Ni)$_3$O$_4$ and its gallium substituted analogs (10\%-Ga, 20\%-Ga, and 40\%-Ga) were synthesized by solid state reaction. The spinel phase was initially confirmed by powder x-ray diffraction. Rietveld refinements of the 0\% and 20\% Ga samples are presented in Figure~\ref{XRD_EDX}(a,b) showing excellent agreement with the cubic spinel phase (space group $Fd\bar{3}m$). Structural characterization for the 10\% and 40\% samples are available in the Supplemental Materials. The refined lattice parameters and adjustable oxygen coordinates as a function of Ga substitution are presented in Table~\ref{SummaryTable}. The cubic spinel structure with the general formula $AB_2$O$_4$, as shown in Figure~\ref{XRD_EDX}(c), has two distinct cation sublattices. The first, indicated in yellow is the tetrahedrally coordinated $A$-sublattice, while the second is the octahedrally coordinated $B$ sublattice, indicated in blue. Independently, these $A$ and $B$ sublattices form, respectively, a diamond and pyrochlore network. From analogy with other known $3d$ transition metal based spinels, one may infer a strong site preference among the different cations but due to the similar atomic form factors of the elements involved, this cannot be resolved by x-ray diffraction. This site selectivity will be a major focus in the sections that follow.

Two minor impurity phases are evident in \textit{only} the 0\% Ga sample, which are zirconia and rock salt phases making up approximately 1\% and 2\% of the sample by volume, respectively. The zirconia phase originates from the degradation of the milling balls during the homogenization process and is not of concern to the magnetic property measurements. The minor rock salt phase has all of its Bragg reflections sitting directly atop spinel main phase peaks and was only detected due to the high resolution of our diffractometer. The refined lattice parameter for the rock salt impurity is in good agreement with that of NiO \cite{rocksaltoxides-structure-nio} and we therefore tentatively assign the likely origin of this phase as being due to Ni segregation, which we will further discuss later in the context of the local environment. 

Energy dispersive x-ray spectroscopy (EDS) measurements, shown for the 0\% and 20\% Ga samples in Figure~\ref{XRD_EDX}(d,e) were employed to investigate the composition and homogeneity of these samples on a microscopic length scale. The grain morphology of the samples is observed to evolve considerably as a function of Ga substitution, despite the identical synthesis method, suggesting it is intrinsic in origin. Likewise, while the elemental mapping is largely homogeneous for the various ions, a higher degree of clustering and inhomogeneity is evident for the Ga-substituted samples. The current results exclude any possible phase segregation on the micron length scale accessible with SEM, and future measurements with a transmission electron microscope (TEM) can help us confirm the homogeneity on the nanometer length scale. The current SEM-EDS mapping and the XRD results support the single phase nature of our spinel HEO samples for different Ga doping levels. Magnetic characterization, which will be shown next, also supports the single phase nature of the magnetism.

\subsection{Magnetic Susceptibility: Onset of ferrimagnetic order}

We begin our magnetic characterization of the spinel HEO by considering the temperature dependent magnetic susceptibility for (Cr,Mn,Fe,Co,Ni)$_3$O$_4$, measured between 2 and 600~K, which is presented in Fig.~\ref{Susceptibility_Summary}(a). The onset of long-range magnetic order at $T_C = 403$~K~\footnote{We define $T_C$ by a slope method wherein the approximately linear increase in the susceptibility is extrapolated to its $x$-intercept.} is marked by a sharp increase in the susceptibility and a bifurcation of the field-cooled and zero field-cooled susceptibilities, measured in an applied field of $H = 100$~Oe. While the low-field susceptibility data could be consistent with either ferro- or ferri-magnetic order, the latter is confirmed via isothermal magnetization measurements, presented in Fig.~\ref{Susceptibility_Summary}(b). At $T=2$~K, the saturated magnetic moment reaches a value of $\mu_{\text{sat}} = 1.52$~$\mu_\text{B}$ per formula unit (FU), which is smaller than even the smallest possible moment for any of the five magnetic transition metals involved and significantly smaller than the average expected moment. Therefore, we can conclude that the ordered moments on the $A$ and $B$ sublattices are aligned antiparallel and the saturated moment measured here represents the difference between the sublattice magnetizations ($\mu_{\text{sat}} = |\mu^A_{\text{ord}} - \mu^B_{\text{ord}}|$). With the exception of the $T=400$~K data set, which is above the ordering temperature, a measurable but small hysteresis is detected at all other temperatures, with the coercive field reaching $H_c = 36$~mT at 2~K, characteristic of soft magnetic behavior. Finally, we note the appearance of a second feature in the zero field-cooled susceptibility at $T^*= 55$~K, which is more evident when viewing the derivative with respect to temperature, plotted in the top panel of Fig.~\ref{Susceptibility_Summary}(a). As this feature only appears in the zero field-cooled data, we tentatively assign it as a depinning transition, where thermal effects overcome disordering allowing some domains to spontaneously realign towards the applied field. Our magnetometry results on (Cr,Mn,Fe,Co,Ni)$_3$O$_4$ are in good agreement with previous reports~\cite{musico2019tunable,sarkar2021comprehensive,cieslak2021magnetic}.

The field-cooled magnetic susceptibility for (Cr,Mn,Fe,Co,Ni)$_3$O$_4$ and each of the Ga-substituted samples is presented in Fig.~\ref{Susceptibility_Summary}(c) and zero field-cooled data are available in the Supporting Information. The susceptibility data show that the onset of long-range magnetic order is suppressed continuously by magnetic dilution via substitution of non-magnetic Ga, reaching $T_C = 121$~K for the 40\%-Ga sample. Meanwhile the $T^*$ depinning transition temperature is found to increase with Ga-concentration (see derivative plots in the Supporting Information), consistent with increasing magnetic disorder, and is altogether absent in the 40\%-Ga sample, apparently coinciding with $T_C$. The evolution of $T_C$ and $T^*$ is presented in Fig.~\ref{Susceptibility_Summary}(e), showing that $T_C$ decreases approximately linearly with Ga-concentration.  

Magnetization isotherms collected at $T=2$~K for each sample are shown in Fig.~\ref{Susceptibility_Summary}(d) while higher temperatures are shown in the Supporting Information. Naively, one might expect the replacement of magnetic transition metals with non-magnetic Ga to yield a monotonic decrease in the net moment, which is not borne out by the data. Instead, the saturated moment, when normalized per formula unit, jumps from $\mu_{\text{sat}} = 1.52$~$\mu_\text{B}$/FU in the 0\% Ga sample to over $2$~$\mu_\text{B}$/FU for both 10\% and 20\% Ga. Only at 40\% Ga does the saturated moment finally drop below the 0\% value. Normalizing the same data per magnetic transition metal, as shown in Fig.~\ref{Susceptibility_Summary}(f), shows a similar trend with the exception of the 40\% sample, which has $\mu_{\text{sat}} = 0.68$~$\mu_\text{B}$/TM that exceeds the 0\% sample. There are two possible contributing factors to these observations: (i) the non-magnetic Ga selectively occupies one of the sublattices yielding a larger difference between the sublattice ordered moments ($\mu_{\text{sat}} = |\mu^A_{\text{ord}} - \mu^B_{\text{ord}}|$) and (ii) in order to satisfy charge neutrality, Ga-substitution induces a change of valence for one or more of the cations from 3+ towards 2+, increasing the average moment per transition metal. Magnetometry data alone do not provide enough information to disentangle these two effects and we will return to both of these points in the section on neutron diffraction and x-ray absorption spectroscopy that follow. 

The $T=2$~K magnetization isotherms also exhibit a pronounced magnetic hardening as a function of Ga concentration. There is a moderate increase in the coercive field going from 0\% to 20\% and a dramatic increase at 40\% Ga, where $H_c = 880$~mT. This effect can be partially attributed to increasing magnetic disorder. As will be shown in the x-ray absorption section, we can account for this observation by considering the occupancy of Co$^{2+}$, which is known to commonly exhibit strong anisotropy in octahedral crystal field environments~\cite{sucksmith1954magnetic,lines1963magnetic}. Indeed, we find that in the 0\% sample with negligible coercivity, the Co$^{2+}$ sits entirely on the tetrahedral site while the increasing coercivity with Ga concentration tracks with the systematic shift of Co$^{2+}$ from the tetrahedral to octahedral site We therefore conclude that through understanding the site selectivity of these HEO spinels, we can likely apply fine-tuning of the elemental composition to optimize magnetic properties.

\subsection{Neutron diffraction: Magnetic structure determination}

\begin{figure*}[!ht]
  \centering
  \includegraphics[width=17cm]{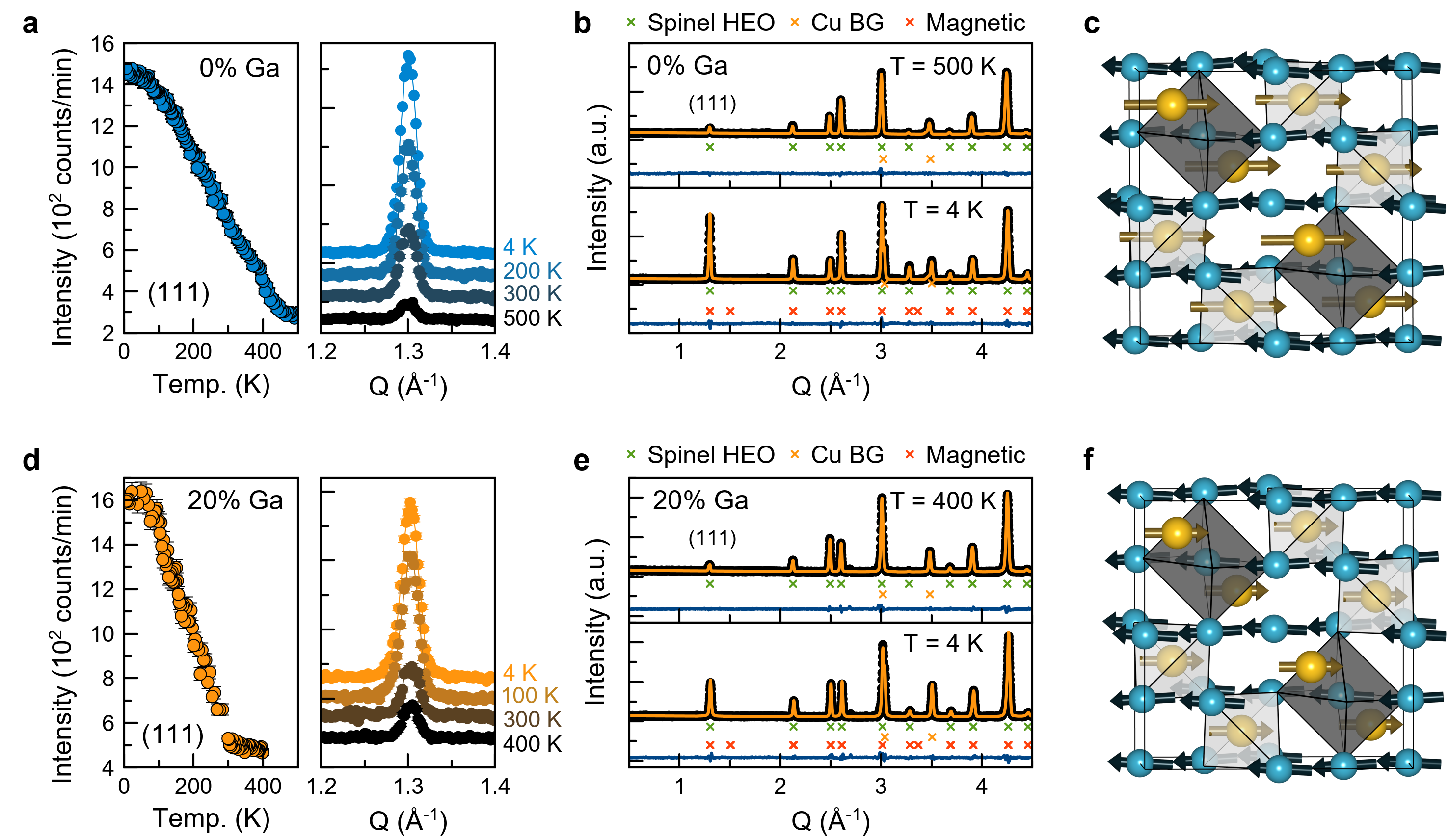}
  \caption{Neutron diffraction measurements and magnetic structure determination for the HEO spinels. The top set of panels show data for the 0\% Ga sample and the bottom set of panels are for the 20\% Ga sample. (\textbf{a,d}) The temperature dependence of the (111) Bragg peak peak shows a sharp increase in intensity indicating the onset of $k=0$ magnetic order. (\textbf{b,e}) Rietveld refinements in the paramagnetic ($T= 500$~K for the 0\% Ga sample and $T=400$~K for the 20\% Ga sample) and magnetically ordered ($T=4$~K) states allow the determination of the magnetic structure. Tick marks indicate the positions of Bragg peaks for the spinel nuclear structure (green), the copper background from the sample environment (yellow), and the $k = 0$ magnetic structure (orange). (\textbf{c,f}) Both the 0\% Ga and 20\% Ga samples exhibit the ferrimagnetic $\Gamma_9$ structure, in which the tetrahedral moments in yellow point exactly along the $\langle100\rangle$ while the octahedral moments in blue are slightly canted away from $\langle100\rangle$. Ga-substitution strongly suppresses the tetrahedral moment.}
  \label{Neutron_Fig}
\end{figure*}

To complete our characterization of the magnetically ordered state in the spinel HEO, we performed powder neutron diffraction measurements on the 0\% and 20\% Ga-substituted samples. There has been one previous neutron diffraction measurement on (Cr,Mn,Fe,Co,Ni)$_3$O$_4$~\cite{sarkar2021comprehensive}; however, this study only includes only a single data set, collected at $T=300$~K, which does not allow the nuclear and magnetic structures to be separately determined, nor does it represent the magnetic ground state since the magnetism continues to evolve considerably below 300~K, as shown by the magnetic susceptibility data in Fig.~\ref{Susceptibility_Summary}(a). In our experiment, we performed measurements between $T=4$ and 500~K, giving us access to both the paramagnetic regime as well as within the ordered state.

Starting with first the 0\% Ga sample, upon cooling through the ordering transition at $T_C = 403$~K, we observe the formation of intense magnetic Bragg peaks that sit atop the nuclear Bragg peaks, as can be seen by viewing the intensity of (111) Bragg peak as a function of temperature shown in Fig.~\ref{Neutron_Fig}(a). The onset of this magnetic order agrees well with the transition temperature detected from susceptibility and we can see that the ordered moment continues to grow monotonically with decreasing temperature before finally saturating at close to 50~K. Comparing the (111) Bragg peak at 500~K, where it is purely nuclear in origin, and 4~K, where it is overwhelmingly magnetic in origin, we can see that there is no significant difference in the line width, indicating that within the resolution limit of this diffractometer, the magnetic order is truly long-range in nature. We can estimate the minimum correlation length by considering the full-width and half maximum of the (111) Bragg peak and this gives $\xi=\sfrac{2\pi}{\text{FWHM}}=294(2)$~\AA.

In order to determine the nature of the magnetically ordered state, we perform a Rietveld analysis of the full diffraction pattern, as shown in Figure~\ref{Neutron_Fig}(b)). Above the onset of magnetic order, at $T=500$~K, our refinement shown in the top panel includes three phases: (i) the cubic spinel structure with space group $Fd\bar{3}m$, (ii) a metallic Cu phase, which originates from the sample environment, and (iii) minor ZrO$_2$ and NiO phases (as described previously in the x-ray diffraction section, tick marks for these phases are omitted for clarity). The coherent scattering lengths of our five transition metals span a large range, from $-3.73$ barns for Mn to $10.3$ barns for Ni. One  might therefore hope to be able to independently refine the site occupations from the neutron data. However, in our analysis, we found that due to trade-offs with other refined parameters (such as the thermal parameters, $B_{iso}$, and the oxygen occupancy), we could not meaningfully distinguish via their $R_{\text{wp}}$ values between a range of plausible cation distributions. Given that our primary goal is to refine the magnetic structure, we therefore constrained the number of adjustable parameters in our refinement of the nuclear structure by allowing only minor deviations from the cation distribution deduced from the XMCD analysis, which will be described in the section that follows, and fixed the oxygen occupancy at 100\%. This yielded excellent agreement between the data and the model and physically reasonable values for all other refined parameters. The temperature dependence of the cubic lattice parameter follows conventional thermal contraction, as shown in the Supporting Information.

\begin{table*}[htbp]
\caption{Structural and magnetic properties of the Ga-substituted HEO spinels determined from x-ray diffraction (cubic lattice parameter $a$ and adjustable oxygen coordinate), magnetic susceptibility (ordering temperature $T_C$ and depinning transition $T^*$), magnetization (saturated moment per formula unit and per transition metal, and coercive field) and magnetic neutron diffraction (tetrahedral and octahedral sublattice moments for 0\% and 20\% samples only).}
\begin{tabular}{lC{2.2cm}C{1.8cm}C{1.5cm}C{1.5cm}C{1.6cm}C{1.6cm}C{1.7cm}C{1.4cm}C{1.4cm}}
  \toprule\multicolumn{1}{c}{} & $a$                  & O coord. & $T_C$ & $T^*$ & $\mu_{\text{sat}}$/FU & $\mu_{\text{sat}}$/TM & Coerc. & $\mu_{\text{tet}}$ & $\mu_{\text{oct}}$ \\
\multicolumn{1}{c}{} & (\AA) &               & (K)   & (K)   & ($\mu_{\text{B}}$)    & ($\mu_{\text{B}}$)    & (mT)   & ($\mu_{\text{B}}$) & ($\mu_{\text{B}}$) \\
\hline
0\% Ga               & 8.3539(3)                & 0.2558(5)     & 382     & 55      & 1.51                                     & 0.50                                     & 36          & 3.60(6)                               & 2.5(1)                                \\
10\% Ga              & 8.3375(2)                     &  0.2580(3)          & 319     & 81      & 2.33                                     & 0.86                                     & 40          & -                                    & -                                    \\
20\% Ga              & 8.3397(2)                & 0.2569(3)     & 230     & 100     & 2.18                                     & 0.91                                     & 98         & 2.46(6)                               & 2.49(9)                               \\
40\% Ga               & 8.3542(2)                    & 0.2566(3)          & 102     & -        & 1.23                                     & 0.68                                     & 880         & -                                  & -\\
\toprule
\end{tabular}
\label{SummaryTable}
\end{table*}

Upon cooling into the magnetically ordered state, we observe that all magnetic Bragg peaks are located at positions allowed by the face-centered cubic selection rules of the nuclear structure. Therefore, we can characterize the magnetic order as having a propagation vector of $k=0$. We next consider the symmetry allowed magnetic structures for the tetrahedral and octahedral sites, which correspond to the $8a$ and $16d$ Wyckoff sites, respectively. Following the representation analysis formalism, we can define these symmetry allowed magnetic structures according to their irreducible representations ($\Gamma$ composed of basis vectors $\psi$). There are two possible $k=0$ states for the tetrahedral site, which are $\Gamma_8$ and $\Gamma_9$, while there are four possible states for the octahedral site: $\Gamma_3$, $\Gamma_5$, $\Gamma_7$, and $\Gamma_9$. In cases where there is a coupled ordering transition involving two sublattices simultaneously, in the case of a second order transition, it is required that both sites order in the same representation group. Thus, by symmetry arguments alone one can deduce that the ordering should involve the $\Gamma_9$ irreducible representations for both the $A$ and $B$ sublattice. However, this assessment can be validated through checking all linear combinations of the irreducible representations and indeed, only a state involving $\Gamma_9$ on both sublattices can give a satisfactory agreement with the data. The resulting refinement at $T=4$~K is shown in the lower panel of Figure~\ref{Neutron_Fig}(b), which in addition to the $\Gamma_9$ magnetic structure contains the same set of structural phases as the $T=500$~K refinement. 

The spin configuration for the refined $\Gamma_9$ magnetic structure of the spinel HEO is shown in Figure~\ref{Neutron_Fig}(c) (magnetic space group \#141.557, $I4_1/am'd'$). The $\Gamma_9$ representation for the tetrahedral site is a ferromagnetic arrangement composed of a single type of basis state (triply degenerate for cubic symmetry) in which the moments points exactly along the cubic $\langle$100$\rangle$ directions with an ordered moment of $\mu_{\text{tet}}=$3.60(6)~$\mu_{\text{B}}$, represented by the yellow sublattice in Fig.~\ref{Neutron_Fig}(c). The $\Gamma_9$ representation for the octahedral site is composed of two types of basis states (each triply degenerate for cubic symmetry) in which the moments point along non-coplanar $\langle$111$\rangle$ and $\langle$211$\rangle$ directions, respectively. Linear combinations of these two basis states can also produce a ferromagnetic arrangement  with moments pointing along the cubic $\langle$100$\rangle$ directions with an ordered moment of $\mu_{\text{oct}}=$2.5(1)~$\mu_{\text{B}}$, as shown for the blue sublattice in Fig.~\ref{Neutron_Fig}(c). The two sublattices are aligned anti-parallel to one another, confirming the ferrimagnetic order deduced from our magnetization data. In our refinement, the best agreement with the experimental data occurred when the moments on the octahedral sites are canted by a small amount, $\alpha = 6.1$ degrees from the $\langle$100$\rangle$ direction. Such canting is well known for ferrimagnetic spinels and is often referred to as the Yafet-Kittel angle~\cite{yafet1952antiferromagnetic,murthy1969yafet}. Detailed characterizations of many spinel solid solutions has shown that the magnitude of this canting is highly correlated with the specific pairwise interactions between the tetrahedral ($A$) and octahedral ($B$) sublattices, which in turn depends strongly on the specific composition and the degree of inversion and magnetic dilution~\cite{willard1999magnetic}. The presence of canting indicates a certain degree of competition between inter- ($J_{AB}$) and intra- ($J_{AA}$ and $J_{BB}$) sublattice exchange interactions, with a strong, antiferromagnetic $J_{BB}$ disrupting the antiferromagnetic coupling between the $A$ and $B$ sublattices. The refined net moment is $\mu_{\text{net}}=$1.4(2)~$\mu_{\text{B}}$ per formula unit, in excellent agreement with the saturated moment observed in magnetization. No change in the magnetic structure is detected at $T^* = 55$~K consistent with our assignment that it is a depinning transition.

For the 20\% Ga substituted sample, we see similar behavior with a sharp increase in the intensity of the (111) Bragg peak that correlates well with $T_C = 284$~K detected by magnetic susceptibility (Fig.~\ref{Neutron_Fig}(d)). Similar to our protocol in the 0\% Ga case, in our Rietveld refinement of the 20\% Ga sample we allow only small deviations from the cation distribution deduced from the XMCD analysis to be shown in the next section. The resulting refinement of the structure in the paramagnetic state at $T = 400$~K gives excellent agreement with the data as shown in Fig.~\ref{Neutron_Fig}(e). The refinement in the magnetically ordered state, at $T=4$~K confirms that the magnetic order is described by the same $\Gamma_9$ ferrimagnetic state observed in the 0\% sample, with an unchanged $B$ sublattice ordered moment of $\mu_{\text{oct}}=$2.49(9)~$\mu_{\text{B}}$ but a substantially reduced $A$ site ordered moment of $\mu_{\text{tet}}=$2.46(6)~$\mu_{\text{B}}$, which is approximately two-thirds the value obtained for the 0\% Ga sample. This supports our previous claim that non-magnetic gallium is preferentially occupying the tetrahedral site. An additional consequence of the highly suppressed $A$ sublattice moment, is that there is a less complete cancellation between the antiparellel $A$ and $B$ sublattice moments such that the net moment increases to $\mu_{\text{net}}=$2.5(2)~$\mu_{\text{B}}$, as compared to $\mu_{\text{net}}=$1.4(2)~$\mu_{\text{B}}$ in the 0\% sample. This also accounts for the observed increase in the saturated moment found in magnetization measurements. Finally, the refined canting angle for the octahedral moments in the 20\% Ga sample, $\alpha=4.6$ degrees, is reduced from the 0\% case, consistent with reduced competition between inter- and intra-sublattice exchange interactions.

\begin{figure*} 
  \centering
  \includegraphics[width=17cm]{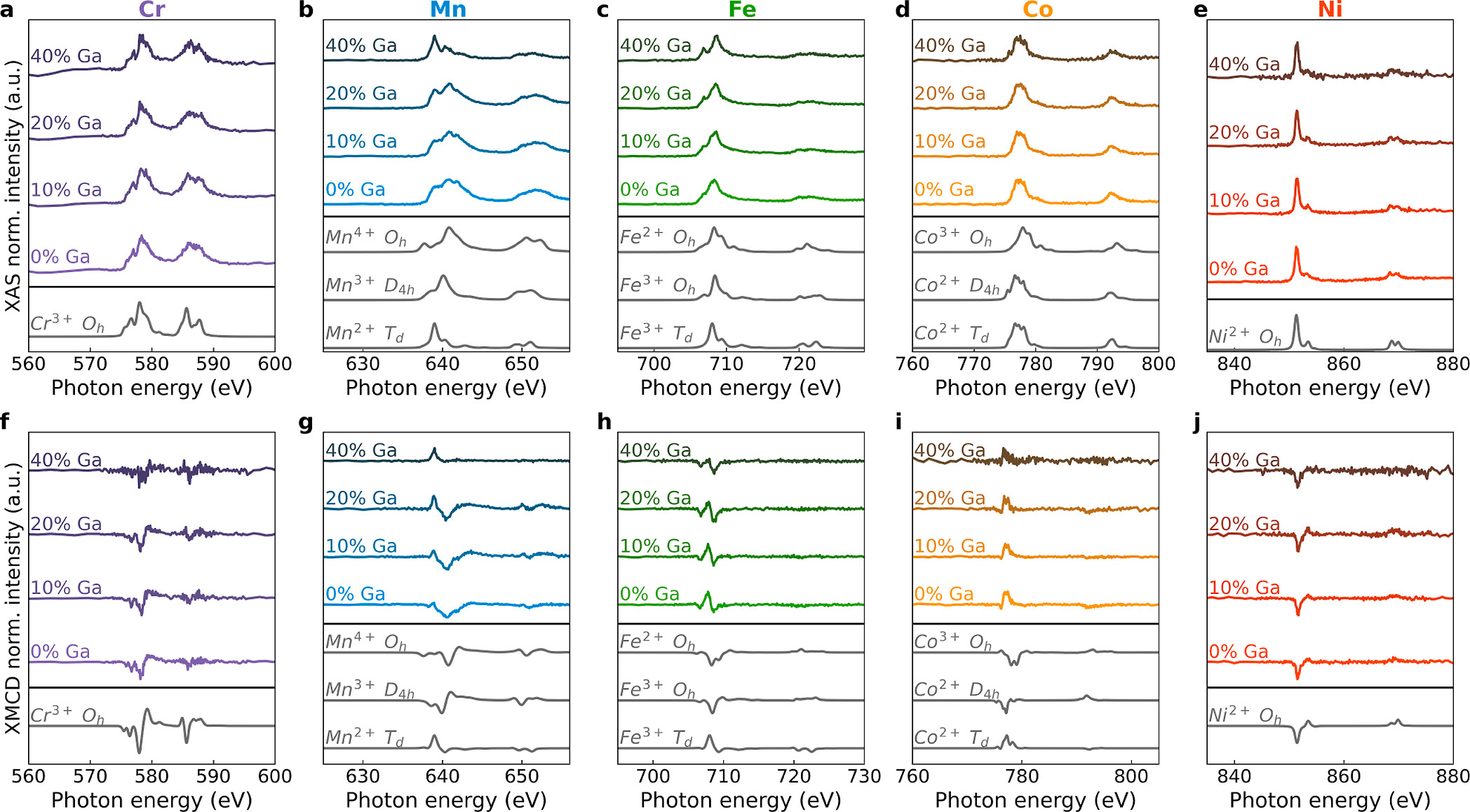}
  \caption{Normalized x-ray magnetic circular dichroism (XMCD) data and multiplet calculations from measurements performed on the HEO spinel samples with varying Ga concentration at the transition metal L\textsubscript{2,3} edges. Top row: L+R spectra for (\textbf{a}) Cr, (\textbf{b}) Mn, (\textbf{c}) Fe, (\textbf{d}) Co, and (\textbf{e}) Ni. The coordination environment can be deduced from these signals, which are approximately equivalent to pure x-ray absorption spectroscopy (XAS) measurements. Bottom row: L-R (XMCD) spectra for (\textbf{f}) Cr, (\textbf{g}) Mn, (\textbf{h}) Fe, (\textbf{i}) Co, and (\textbf{j}) Ni. The sign of each spectrum reflects the ferrimagnetic arrangement in the HEO spinels. Predominantly positive signals correspond to mainly tetrahedral states (T$_d$) while negative signals indicate a mostly octahedral coordination (O$_h$).}
  \label{HEO_multiplet}
\end{figure*}

\subsection{XAS and XMCD: Local ordering and site selectivity}

X-ray magnetic circular dichroism (XMCD) is one of the select few pathways to determining site occupancies in transition-metal spinel oxides \cite{Pattrick-VanDerLaan-cation-occupancy-spinels} due to its simultaneous chemical and crystal field sensitivity. Site occupancies can be deduced by comparison with multiplet calculations, which depend on both the coordination and the valence of the transition metal in question. XMCD measurements for the spinel HEO samples, along with the corresponding multiplet calculations, are presented in Fig.~\ref{HEO_multiplet}, where the top row of panels are the sum of the left and right circularly polarized responses (L+R) and therefore correspond approximately to x-ray absorption spectroscopy (XAS) while the bottom row is the difference (L-R), corresponding to the dichroism. The spectra are presented normalized to their maxima, so the main features can be resolved across the entire composition range. An important feature of these spectra is that, for ferrimagnetic spinels, one can observe the antiferromagnetic coupling of the tetrahedral and octahedral sites in a straightforward manner. The sign of the XMCD signal is directly dependent on the number of available holes in the transition metal ions' $d$-shell and since the holes for the octahedral and tetrahedral sites have opposing spin, the XMCD will have opposite sign. While it is not possible to directly measure the site occupancy of Ga in this way due to its filled $d$-shell, its distribution can be approximately deduced from the constraint that the stoichiometry respects the spinel structure. As analysis of dichroism requires the sample to be in a field polarized state, an important limitation occurs for the 40\% Ga sample, which is not close to saturation in our experiment due to its large coercive field. For this reason, the XMCD of the 40\% Ga sample represents a semi-random arrangement of magnetic domains that cannot be meaningfully interpreted and this is also why the observed signal to noise in this sample is significantly higher. Our analysis of the 40\% Ga sample therefore relies on only the conventional absorption data.

The analysis of Cr and Ni from Fig.~\ref{HEO_multiplet}(a,f) and (e,j) is straightforward as neither are observed to change as a function of Ga concentration. Both are found exclusively in an octahedral (O$_h$) coordination, as indicated by the negative dichroism with Cr in a 3+ oxidation state and Ni in a 2+ oxidation state. In both cases, the multiplet calculation describes well all the features in both the absorption and the dichroism spectra. The remaining cations will be discussed in order of increasing complexity. 

The next ion we consider is Fe, which can be seen in Fig.~\ref{HEO_multiplet}(c). For all compositions, we observe a double peak structure in the XAS that cannot be described by any one of the possible multiplet calculations individually; instead Fe exists in an admixture of coordinations in all samples, requiring a more nuanced analysis. As a function of increasing Ga concentration, we can observe that the XAS signal becomes progressively sharper, with the 40\% Ga concentration sample exhibiting a sharp peak at 707~eV that aligns well with the multiplet calculation for Fe$^{3+}$ in octahedral coordination. Simultaneously, the predominantly positive XMCD signature in the 0\% Ga sample, which is consistent with Fe$^{3+}$ in tetrahedral coordination, evolves towards a larger negative contribution with increasing Ga, suggesting a general trend of a Fe shifting from predominantly tetrahedral towards predominantly octahedral coordination. We therefore conclude that Fe remains in a 3+ oxidation state  across the analyzed composition range and  gradually shifts from mainly tetrahedral at 0\% towards a primarily octahedral state around 20\% Ga substitution.

Considering next Co, we can observe that the XAS spectra for all Ga concentrations appear at first largely similar but a closer examination reveals the presence of a pronounced shoulder-like feature on the lower energy side of the central peak in only the 40\% Ga concentration sample  (Fig.~\ref{HEO_multiplet}(d)). Comparison of this data set with the multiplet calculations reveals that the shoulder-like feature in the 40\% Ga sample originates from a majority composition of Co$^{2+}$ in a tetragonally-distorted octahedron (D$_{4h}$), originating from a Jahn-Teller effect. This distortion is required to properly reproduce the relative intensities and features in the pre and post-edge of the 40\% XAS, and is known to occur in Co$^{2+}$ oxides and compounds \cite{Multiplet-Co-DeGroot,Co-2-JT-ref}. Upon closer inspection of the spectra corresponding to 10\% and 20\% Ga-diluted samples, the emergence of these features can already be observed, albeit on a smaller scale. However, the XMCD spectra show conclusively that for the lower Ga concentrations, the majority species is Co$^{2+}$ in a tetrahedral coordination. The XMCD for the 40\% Ga sample shows a near complete cancellation of the signal but we must stress again that the XMCD of the 40\% Ga sample is not reliable due to the sample not being in its field saturated state and we therefore must base our conclusions on the XAS and other supporting data. Detailed fitting of the XAS data is presented in in the Supplemental Information. Important corroboration for the shift of Co$^{2+}$ from the tetrahedral to the octahedral site in the 40\% Ga sample comes from the magnetization. Co$^{2+}$ in this distorted octahedral environment also elucidates the origin of the increased coercivity observed in Fig.~\ref{Susceptibility_Summary}(d). While more subtle than the effect it has on other ions, this change is then identified as a direct consequence of Ga substitution.

\begin{figure*}
  \centering
  \includegraphics[width=17cm]{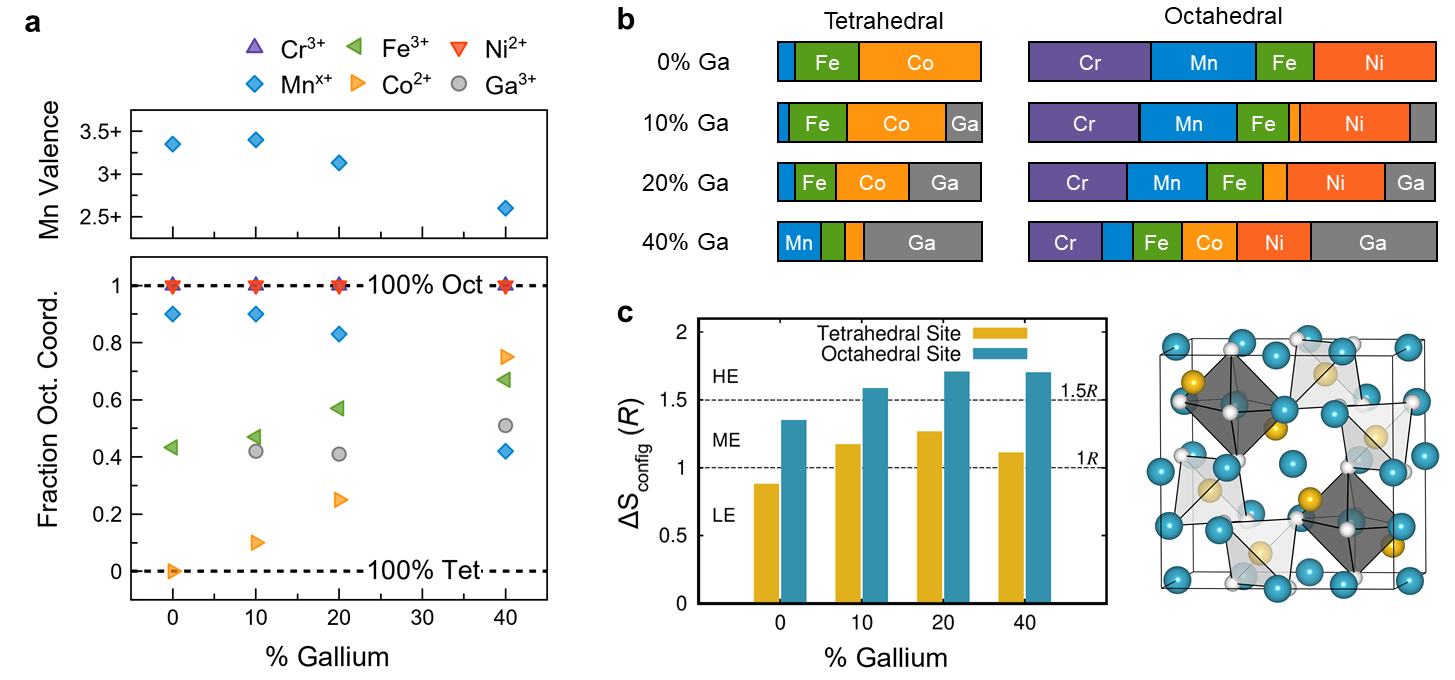}
  \caption{Summary of cation distributions and sublattice entropies in the spinel HEO as a function of Ga substitution. (\textbf{a}) The valence of Mn and the fraction of octahedral occupation as a function of Ga concentration. Preferential occupation of Ga onto tetrahedral sites causes a shift of Fe and Co towards the octahedral site and the stable 3+ valence of Ga is compensated by a reduction in the valence of Mn, which in its 2+ valence state prefers tetrahedral sites. (\textbf{b}) Schematic representation of the cation occupation of the tetrahedral and octahedral sites for each sample. (\textbf{c}) Calculated sublattice entropy assuming ideal configurational disorder as a function of Ga concentration for the tetrahedral site in yellow and the octahedral site in blue. At 0\% Ga, neither sublattice surpasses the empirical threshold for high entropy but with Ga substitution high and medium entropy is achieved for the octahedral and tetrahedral sites, respectively.}
  \label{Summary_Fig}
\end{figure*}

The final ion to consider is Mn, whose behavior is the most challenging to explain and reproduce with multiplet calculations, particularly for the low Ga-concentration samples. From Fig.~\ref{HEO_multiplet}(b), we observe three broad features centered around 640 eV in the L$_3$ edge at lower Ga substitution, which are replaced by two comparatively sharp features at 40\% Ga concentration. The features in the 40\% sample are well explained by the spectrum for Mn$^{2+}$ in tetrahedral coordination. As a function of decreasing Ga concentration, the Mn$^{2+}$ tetrahedral signal persists but becomes weaker as simultaneously the XMCD becomes increasingly negative. At 0\% the most prominent signal is located at 640.7 eV and this feature is attributed to Mn$^{3+}$ in a tetragonally distorted octahedron (D$_{4h}$ symmetry) due to a strong Jahn-Teller effect. The final feature in the L$_3$ edge is located at 641.9 eV, and corresponds to the presence of Mn$^{4+}$ in octahedral coordination. The evolution of Mn in the sample can therefore be characterized as follows: the undiluted sample is characterized by the presence of all three Mn species, 2+ tetrahedral, 3+ tetragonally distorted octahedral, and 4+ octahedral. Upon the introduction of Ga to the structure, Mn accommodates the extra charge by gradually shifting towards the Mn$^{2+}$ state in tetrahedral coordination, as shown in the top panel of \ref{Summary_Fig}(a). At 20\% Ga, the Mn$^{4+}$ octahedral signal has been reduced substantially, and by 40\% Ga, Mn is predominantly in a 2+ tetrahedral state, with minimal 3+ D$_{4h}$ and 4+ octahedral contributions.

Putting the analysis of all XMCD results together allows us to arrive at approximate site distributions and compositions for each of our four samples, as summarized in Fig.~\ref{Summary_Fig}(a,b). We can see that the undiluted 0\% Ga sample presents a significant degree of site selectivity, with Cr and Ni exclusively occupying octahedral sites, Co only in tetrahedral sites, and mixed occupations for Mn and Fe, leading to the estimated composition for the tetrahedral site of $A =$~Mn$^{2+}_{0.06}$Fe$^{3+}_{0.33}$Co$^{2+}_{0.61}$ and for the octahedral site of $B =$~Cr$^{3+}_{0.61}$Mn$^{3+}_{0.28}$Mn$^{4+}_{0.27}$Fe$^{3+}_{0.28}$Ni$^{2+}_{0.54}$, which takes into account the deficiency in Ni due to the minor rock salt impurity phase described previously. The result for the 0\% Ga sample can be compared with the previous study by Sarkar \textit{et al.}, where it was concluded that Co was present in an almost equal mixture of 3+ and 2+ tetrahedral states, while Mn was ascribed as being almost entirely 3+ octahedral with a minor 2+ octahedral component~\cite{sarkar2021comprehensive}. One factor that might account for some of these differences is that the data collected in Ref.~\cite{sarkar2021comprehensive} used a total electron yield method, which is more surface sensitive as compared to the inverse partial fluorescence yield method used here. Another intriguing possibility is that the fundamental differences between the results could be indicative of some dependence of the site selectivity on the synthesis method, choice of reagents, or thermal treatments performed on the samples. Nevertheless, it is interesting to note that despite the differences in oxidation state, the elemental composition of the octahedral and tetrahedral sites found here agrees almost exactly with that of Ref.~\cite{sarkar2021comprehensive}.

Upon the introduction of Ga into the HEO spinel, the compositions of each sample can be qualitatively estimated from the XMCD spectra, with the distribution of Ga ions in each sample constrained so that the final composition respects the stoichiometry of the spinel structure. While Ga is divided almost evenly between the tetrahedral and octahedral sites (Fig.~\ref{Summary_Fig}(a)), due to the 1:2 sublattice ratio in the spinel structure a larger fraction of the tetrahedral magnetic ions are replaced. Therefore, we find a preferential dilution of the tetrahedral sublattice, as shown schematically in Fig.~\ref{Summary_Fig}(b), consistent with the trends observed in our magnetization data. Meanwhile, Co and Fe accommodate by gradually shifting to primarily octahedral crystal field environments (Fig.~\ref{Summary_Fig}(a)). On the other hand, Mn seems to adapt to the additional charge (Ga's 3+ oxidation state is larger than the 2.667+ average for the spinel structure) by reducing towards a 2+ oxidation state, as shown in the upper panel of Fig.~\ref{Summary_Fig}(a). At 20\% Ga substitution, we find that the tetrahedral site has been 35\% diluted by non-magnetic Ga while the octahedral site has only been diluted by just over 10\%, consistent with the large reduction in the tetrahedral moment observed in our neutron diffraction analysis. The negligible change in the octahedral sublattice moment can be ascribed to the larger fraction of high spin Mn$^{3+}$ as compared to Mn$^{4+}$ at this composition. By 40\% Ga, the ion distribution reaches the estimated composition for the tetrahedral site of $A =$~Mn$^{2+}_{0.21}$Fe$^{3+}_{0.12}$Co$^{2+}_{0.09}$Ga$^{3+}_{0.58}$ and for the octahedral site of $B =$~Cr$^{3+}_{0.36}$Mn$^{3+}_{0.09}$Mn$^{4+}_{0.06}$Fe$^{3+}_{0.24}$Co$^{2+}_{0.27}$Ni$^{2+}_{0.36}$Ga$^{3+}_{0.62}$. Here we can see that non-magnetic Ga occupies more than 50\% of the tetrahedral sites while only occupying approximately 25\% of the octahedral sites. 

\subsection{Entropy analysis}

With these results in mind, we can begin to construct a realistic estimate of the configurational entropy for each of these spinel HEOs taking into account the preferential occupation of multiple cation sublattices. Our deduced compositions, which are shown schematically in Fig.~\ref{Summary_Fig}(b) respect the stoichiometry of the spinel structure and also obey charge neutrality with the oxygen sublattice within negligible error. Our assessment of the transition metal ion sites and valences agree with both the experimental XMCD spectra and also align well with the well-established crystal field behaviors for these ions. From this estimate, one can then compute the corresponding configurational entropy,
\begin{equation}
    S_{\rm{config}}= -R\sum_{ions} x\ln{x}
\end{equation}
where $x$ is the fractional occupation of each ion inhabiting the site. Empirically it has been established that the threshold for medium entropy occurs at $S_{\rm{config}}>1R$ and high entropy occurs at $S_{\rm{config}}>1.5R$ where the conventional 5-component equiatomic HEO has $S_{\rm{config}}=1.61R$. The computed configurational entropy for each of our HEO spinels, separated by sublattice, is plotted in Fig.~\ref{Summary_Fig}(c). One important limitation of such a naive calculation is that it does not take into account any level of clustering or pairwise atomic correlations. While our Ga-substituted samples are single phase, elemental mapping of these materials shown previously does suggest some level of micron scale inhomogeneity that places a limit on the true configurational entropy. Therefore the entropy figures shown in Fig.~\ref{Summary_Fig}(c) should be considered as a strict upper bound.

One can deduce on general grounds that site selectivity in the spinel structure is a force that will act to reduce configurational entropy. Indeed in the 0\% Ga case we find that the tetrahedral site, which is majority occupied by Co, falls in the low-entropy regime with $S_{\rm{config,tet}}=0.9R$, while the octahedral site has medium-entropy $S_{\rm{config,oct}}=1.35R$. However, if we consider the two different valences of Mn as contributing to the configurational entropy, this alone further enhances the octahedral to the high-entropy limit with $S_{\rm{config,oct}}=1.52R$. Likewise, the introduction of Ga as a 6th cation has an overall enhancing effect on the configurational entropy, with the tetrahedral site being elevated to medium-entropy and the octahedral site being raised to the high-entropy limit by 10\% Ga. The maximum combined configurational entropy, among compositions we've studied is reached for the 20\% Ga sample, which exceeds $S_{\rm{config,total}}=3R$. This finding implies that, with appropriate cation choice, the configurational entropy of the two sublattices of the spinel structure can likely be independently tuned and precisely controlled across different entropy regimes, which we term ``entropy engineering''.

\section{Conclusions}

In this work, we have demonstrated that due to intense site selectivity, the magnetic ground state of the spinel HEO can be strongly tuned by nonmagnetic Ga substitution. Magnetic susceptibility measurements show that the suppression in the onset of long-range magnetic order and  simultaneously, there is a significant enhancement in the coercive field and a nonmonotonic variation in the saturated moments. Our neutron diffraction measurements show that the ferrimagnetic order is robust in response to  Ga-substitution, with a rapidly suppressed tetrahedral moment and a reduced inter sublattice frustration. Finally,  our XMCD measurements reveal that the inclusion of Ga is accommodated through valence and site re-organizations for Mn, Fe, and Co, while Ni and Cr are unaffected. An extremely high entropy system with $S_{\rm{config,total}}>3R$ is realized for the 20\% Ga sample and we show that the individual sublattice entropies can be precisely engineered by the choice of cations. We also demonstrate that multiple valences are a route to further enhancing the configurational entropy.

In light of these findings, many new avenues of inquiry beckon. For instance, are there any subsets of these five magnetic transition metals or different stoichiometries that would yield a non-ferrimagnetic ordered state? What is the fate of the ferrimagnetic order at higher nonmagnetic dilutions? Exploring the limit in which long-range magnetic order is completely suppressed will allow a direct comparison with percolation theory. Along a different axis, the spectral XMCD differences between our work and Sarkar \textit{et al.}~\cite{sarkar2021comprehensive} suggest that there may be significant sample dependence related to synthesis procedure and this topic remains largely unexplored. A better understanding of the exact level of configurational entropy in the spinel HEO is also, for now, out of reach. While our study suggests a very high level of site selectivity, there may be additional preferred short range configurations that will require highly sophisticated studies of ions in a pairwise fashion. We therefore conclude on the optimistic note that the field of high entropy oxides is still in its infancy and many important discoveries remain to be made. 

\section*{Supporting Information}

Experimental details, supporting figures for x-ray diffraction, energy dispersive x-ray spectroscopy, magnetic susceptibility, and neutron diffraction, parameters used in the multiplet theory calculations for the x-ray absorption analysis

\begin{acknowledgments}

The authors thank Joerg Rottler and Solveig Stubmo Aamlid for insightful conversations on high entropy oxides and Jacob Kabel for assistance collecting the energy dispersive x-ray spectroscopy data. This work was supported by the Natural Sciences and Engineering Research Council of Canada and the CIFAR Azrieli Global Scholars program. This research was undertaken thanks in part to funding from the Canada First Research Excellence Fund, Quantum Materials and Future Technologies Program. Part of research described in this paper was performed at the Canadian Light Source, a national research facility of the University of Saskatchewan, which is funded by the Canada Foundation for Innovation, the NSERC, the National Research Council Canada, the Canadian Institutes of Health Research, the Government of Saskatchewan, and the University of Saskatchewan. A portion of this research used resources at the High Flux Isotope Reactor, a DOE Office of Science User Facility operated by the Oak Ridge National Laboratory.

\end{acknowledgments}


\bibliography{bibliography}

\begin{thebibliography}{49}%
\makeatletter
\providecommand \@ifxundefined [1]{%
 \@ifx{#1\undefined}
}%
\providecommand \@ifnum [1]{%
 \ifnum #1\expandafter \@firstoftwo
 \else \expandafter \@secondoftwo
 \fi
}%
\providecommand \@ifx [1]{%
 \ifx #1\expandafter \@firstoftwo
 \else \expandafter \@secondoftwo
 \fi
}%
\providecommand \natexlab [1]{#1}%
\providecommand \enquote  [1]{``#1''}%
\providecommand \bibnamefont  [1]{#1}%
\providecommand \bibfnamefont [1]{#1}%
\providecommand \citenamefont [1]{#1}%
\providecommand \href@noop [0]{\@secondoftwo}%
\providecommand \href [0]{\begingroup \@sanitize@url \@href}%
\providecommand \@href[1]{\@@startlink{#1}\@@href}%
\providecommand \@@href[1]{\endgroup#1\@@endlink}%
\providecommand \@sanitize@url [0]{\catcode `\\12\catcode `\$12\catcode
  `\&12\catcode `\#12\catcode `\^12\catcode `\_12\catcode `\%12\relax}%
\providecommand \@@startlink[1]{}%
\providecommand \@@endlink[0]{}%
\providecommand \url  [0]{\begingroup\@sanitize@url \@url }%
\providecommand \@url [1]{\endgroup\@href {#1}{\urlprefix }}%
\providecommand \urlprefix  [0]{URL }%
\providecommand \Eprint [0]{\href }%
\providecommand \doibase [0]{http://dx.doi.org/}%
\providecommand \selectlanguage [0]{\@gobble}%
\providecommand \bibinfo  [0]{\@secondoftwo}%
\providecommand \bibfield  [0]{\@secondoftwo}%
\providecommand \translation [1]{[#1]}%
\providecommand \BibitemOpen [0]{}%
\providecommand \bibitemStop [0]{}%
\providecommand \bibitemNoStop [0]{.\EOS\space}%
\providecommand \EOS [0]{\spacefactor3000\relax}%
\providecommand \BibitemShut  [1]{\csname bibitem#1\endcsname}%
\let\auto@bib@innerbib\@empty
\bibitem [{\citenamefont {Rost}\ \emph {et~al.}(2015)\citenamefont {Rost},
  \citenamefont {Sachet}, \citenamefont {Borman}, \citenamefont {Moballegh},
  \citenamefont {Dickey}, \citenamefont {Hou}, \citenamefont {Jones},
  \citenamefont {Curtarolo},\ and\ \citenamefont {Maria}}]{rost2015entropy}%
  \BibitemOpen
  \bibfield  {author} {\bibinfo {author} {\bibfnamefont {Christina~M}\
  \bibnamefont {Rost}}, \bibinfo {author} {\bibfnamefont {Edward}\ \bibnamefont
  {Sachet}}, \bibinfo {author} {\bibfnamefont {Trent}\ \bibnamefont {Borman}},
  \bibinfo {author} {\bibfnamefont {Ali}\ \bibnamefont {Moballegh}}, \bibinfo
  {author} {\bibfnamefont {Elizabeth~C}\ \bibnamefont {Dickey}}, \bibinfo
  {author} {\bibfnamefont {Dong}\ \bibnamefont {Hou}}, \bibinfo {author}
  {\bibfnamefont {Jacob~L}\ \bibnamefont {Jones}}, \bibinfo {author}
  {\bibfnamefont {Stefano}\ \bibnamefont {Curtarolo}}, \ and\ \bibinfo {author}
  {\bibfnamefont {Jon-Paul}\ \bibnamefont {Maria}},\ }\bibfield  {title}
  {\enquote {\bibinfo {title} {Entropy-stabilized oxides},}\ }\href
  {https://www.nature.com/articles/ncomms9485?origin=ppub} {\bibfield
  {journal} {\bibinfo  {journal} {Nature communications}\ }\textbf {\bibinfo
  {volume} {6}},\ \bibinfo {pages} {1--8} (\bibinfo {year} {2015})}\BibitemShut
  {NoStop}%
\bibitem [{\citenamefont {Zhang}\ and\ \citenamefont
  {Reece}(2019)}]{zhang2019review}%
  \BibitemOpen
  \bibfield  {author} {\bibinfo {author} {\bibfnamefont {Rui-Zhi}\ \bibnamefont
  {Zhang}}\ and\ \bibinfo {author} {\bibfnamefont {Michael~J}\ \bibnamefont
  {Reece}},\ }\bibfield  {title} {\enquote {\bibinfo {title} {Review of high
  entropy ceramics: design, synthesis, structure and properties},}\ }\href
  {https://doi.org/10.1039/C9TA05698J} {\bibfield  {journal} {\bibinfo
  {journal} {Journal of Materials Chemistry A}\ }\textbf {\bibinfo {volume}
  {7}},\ \bibinfo {pages} {22148--22162} (\bibinfo {year} {2019})}\BibitemShut
  {NoStop}%
\bibitem [{\citenamefont {Oses}\ \emph {et~al.}(2020)\citenamefont {Oses},
  \citenamefont {Toher},\ and\ \citenamefont {Curtarolo}}]{oses2020high}%
  \BibitemOpen
  \bibfield  {author} {\bibinfo {author} {\bibfnamefont {Corey}\ \bibnamefont
  {Oses}}, \bibinfo {author} {\bibfnamefont {Cormac}\ \bibnamefont {Toher}}, \
  and\ \bibinfo {author} {\bibfnamefont {Stefano}\ \bibnamefont {Curtarolo}},\
  }\bibfield  {title} {\enquote {\bibinfo {title} {High-entropy ceramics},}\
  }\href {https://www.nature.com/articles/s41578-019-0170-8} {\bibfield
  {journal} {\bibinfo  {journal} {Nature Reviews Materials}\ }\textbf {\bibinfo
  {volume} {5}},\ \bibinfo {pages} {295--309} (\bibinfo {year}
  {2020})}\BibitemShut {NoStop}%
\bibitem [{\citenamefont {Music{\'o}}\ \emph {et~al.}(2020)\citenamefont
  {Music{\'o}}, \citenamefont {Gilbert}, \citenamefont {Ward}, \citenamefont
  {Page}, \citenamefont {George}, \citenamefont {Yan}, \citenamefont
  {Mandrus},\ and\ \citenamefont {Keppens}}]{musico2020emergent}%
  \BibitemOpen
  \bibfield  {author} {\bibinfo {author} {\bibfnamefont {Brianna~L}\
  \bibnamefont {Music{\'o}}}, \bibinfo {author} {\bibfnamefont {Dustin}\
  \bibnamefont {Gilbert}}, \bibinfo {author} {\bibfnamefont {Thomas~Zac}\
  \bibnamefont {Ward}}, \bibinfo {author} {\bibfnamefont {Katharine}\
  \bibnamefont {Page}}, \bibinfo {author} {\bibfnamefont {Easo}\ \bibnamefont
  {George}}, \bibinfo {author} {\bibfnamefont {Jiaqiang}\ \bibnamefont {Yan}},
  \bibinfo {author} {\bibfnamefont {David}\ \bibnamefont {Mandrus}}, \ and\
  \bibinfo {author} {\bibfnamefont {Veerle}\ \bibnamefont {Keppens}},\
  }\bibfield  {title} {\enquote {\bibinfo {title} {The emergent field of high
  entropy oxides: Design, prospects, challenges, and opportunities for
  tailoring material properties},}\ }\href
  {https://aip.scitation.org/doi/full/10.1063/5.0003149} {\bibfield  {journal}
  {\bibinfo  {journal} {APL Materials}\ }\textbf {\bibinfo {volume} {8}},\
  \bibinfo {pages} {040912} (\bibinfo {year} {2020})}\BibitemShut {NoStop}%
\bibitem [{\citenamefont {Sarkar}\ \emph {et~al.}(2020)\citenamefont {Sarkar},
  \citenamefont {Breitung},\ and\ \citenamefont {Hahn}}]{sarkar2020high}%
  \BibitemOpen
  \bibfield  {author} {\bibinfo {author} {\bibfnamefont {Abhishek}\
  \bibnamefont {Sarkar}}, \bibinfo {author} {\bibfnamefont {Ben}\ \bibnamefont
  {Breitung}}, \ and\ \bibinfo {author} {\bibfnamefont {Horst}\ \bibnamefont
  {Hahn}},\ }\bibfield  {title} {\enquote {\bibinfo {title} {High entropy
  oxides: The role of entropy, enthalpy and synergy},}\ }\href
  {https://doi.org/10.1016/j.scriptamat.2020.05.019} {\bibfield  {journal}
  {\bibinfo  {journal} {Scripta Materialia}\ }\textbf {\bibinfo {volume}
  {187}},\ \bibinfo {pages} {43--48} (\bibinfo {year} {2020})}\BibitemShut
  {NoStop}%
\bibitem [{\citenamefont {McCormack}\ and\ \citenamefont
  {Navrotsky}(2021)}]{mccormack2021thermodynamics}%
  \BibitemOpen
  \bibfield  {author} {\bibinfo {author} {\bibfnamefont {Scott~J}\ \bibnamefont
  {McCormack}}\ and\ \bibinfo {author} {\bibfnamefont {Alexandra}\ \bibnamefont
  {Navrotsky}},\ }\bibfield  {title} {\enquote {\bibinfo {title}
  {Thermodynamics of high entropy oxides},}\ }\href
  {https://doi.org/10.1016/j.actamat.2020.10.043} {\bibfield  {journal}
  {\bibinfo  {journal} {Acta Materialia}\ }\textbf {\bibinfo {volume} {202}},\
  \bibinfo {pages} {1--21} (\bibinfo {year} {2021})}\BibitemShut {NoStop}%
\bibitem [{\citenamefont {B{\'e}rardan}\ \emph {et~al.}(2016)\citenamefont
  {B{\'e}rardan}, \citenamefont {Franger}, \citenamefont {Meena},\ and\
  \citenamefont {Dragoe}}]{berardan2016room}%
  \BibitemOpen
  \bibfield  {author} {\bibinfo {author} {\bibfnamefont {D}~\bibnamefont
  {B{\'e}rardan}}, \bibinfo {author} {\bibfnamefont {S}~\bibnamefont
  {Franger}}, \bibinfo {author} {\bibfnamefont {AK}~\bibnamefont {Meena}}, \
  and\ \bibinfo {author} {\bibfnamefont {N}~\bibnamefont {Dragoe}},\ }\bibfield
   {title} {\enquote {\bibinfo {title} {Room temperature lithium superionic
  conductivity in high entropy oxides},}\ }\href
  {https://pubs.rsc.org/en/content/articlehtml/2016/ta/c6ta03249d} {\bibfield
  {journal} {\bibinfo  {journal} {Journal of Materials Chemistry A}\ }\textbf
  {\bibinfo {volume} {4}},\ \bibinfo {pages} {9536--9541} (\bibinfo {year}
  {2016})}\BibitemShut {NoStop}%
\bibitem [{\citenamefont {Sarkar}\ \emph {et~al.}(2018)\citenamefont {Sarkar},
  \citenamefont {Velasco}, \citenamefont {Wang}, \citenamefont {Wang},
  \citenamefont {Talasila}, \citenamefont {de~Biasi}, \citenamefont
  {K{\"u}bel}, \citenamefont {Brezesinski}, \citenamefont {Bhattacharya},
  \citenamefont {Hahn} \emph {et~al.}}]{sarkar2018high}%
  \BibitemOpen
  \bibfield  {author} {\bibinfo {author} {\bibfnamefont {Abhishek}\
  \bibnamefont {Sarkar}}, \bibinfo {author} {\bibfnamefont {Leonardo}\
  \bibnamefont {Velasco}}, \bibinfo {author} {\bibfnamefont {Di}~\bibnamefont
  {Wang}}, \bibinfo {author} {\bibfnamefont {Qingsong}\ \bibnamefont {Wang}},
  \bibinfo {author} {\bibfnamefont {Gopichand}\ \bibnamefont {Talasila}},
  \bibinfo {author} {\bibfnamefont {Lea}\ \bibnamefont {de~Biasi}}, \bibinfo
  {author} {\bibfnamefont {Christian}\ \bibnamefont {K{\"u}bel}}, \bibinfo
  {author} {\bibfnamefont {Torsten}\ \bibnamefont {Brezesinski}}, \bibinfo
  {author} {\bibfnamefont {Subramshu~S}\ \bibnamefont {Bhattacharya}}, \bibinfo
  {author} {\bibfnamefont {Horst}\ \bibnamefont {Hahn}},  \emph {et~al.},\
  }\bibfield  {title} {\enquote {\bibinfo {title} {High entropy oxides for
  reversible energy storage},}\ }\href
  {https://www.nature.com/articles/s41467-018-05774-5} {\bibfield  {journal}
  {\bibinfo  {journal} {Nature communications}\ }\textbf {\bibinfo {volume}
  {9}},\ \bibinfo {pages} {1--9} (\bibinfo {year} {2018})}\BibitemShut
  {NoStop}%
\bibitem [{\citenamefont {Sarkar}\ \emph {et~al.}(2019)\citenamefont {Sarkar},
  \citenamefont {Wang}, \citenamefont {Schiele}, \citenamefont {Chellali},
  \citenamefont {Bhattacharya}, \citenamefont {Wang}, \citenamefont
  {Brezesinski}, \citenamefont {Hahn}, \citenamefont {Velasco},\ and\
  \citenamefont {Breitung}}]{sarkar2019high}%
  \BibitemOpen
  \bibfield  {author} {\bibinfo {author} {\bibfnamefont {Abhishek}\
  \bibnamefont {Sarkar}}, \bibinfo {author} {\bibfnamefont {Qingsong}\
  \bibnamefont {Wang}}, \bibinfo {author} {\bibfnamefont {Alexander}\
  \bibnamefont {Schiele}}, \bibinfo {author} {\bibfnamefont {Mohammed~Reda}\
  \bibnamefont {Chellali}}, \bibinfo {author} {\bibfnamefont {Subramshu~S}\
  \bibnamefont {Bhattacharya}}, \bibinfo {author} {\bibfnamefont
  {Di}~\bibnamefont {Wang}}, \bibinfo {author} {\bibfnamefont {Torsten}\
  \bibnamefont {Brezesinski}}, \bibinfo {author} {\bibfnamefont {Horst}\
  \bibnamefont {Hahn}}, \bibinfo {author} {\bibfnamefont {Leonardo}\
  \bibnamefont {Velasco}}, \ and\ \bibinfo {author} {\bibfnamefont {Ben}\
  \bibnamefont {Breitung}},\ }\bibfield  {title} {\enquote {\bibinfo {title}
  {High-entropy oxides: fundamental aspects and electrochemical properties},}\
  }\href {https://doi.org/10.1002/adma.201806236} {\bibfield  {journal}
  {\bibinfo  {journal} {Advanced Materials}\ }\textbf {\bibinfo {volume}
  {31}},\ \bibinfo {pages} {1806236} (\bibinfo {year} {2019})}\BibitemShut
  {NoStop}%
\bibitem [{\citenamefont {Sun}\ and\ \citenamefont {Dai}(2021)}]{sun2021high}%
  \BibitemOpen
  \bibfield  {author} {\bibinfo {author} {\bibfnamefont {Yifan}\ \bibnamefont
  {Sun}}\ and\ \bibinfo {author} {\bibfnamefont {Sheng}\ \bibnamefont {Dai}},\
  }\bibfield  {title} {\enquote {\bibinfo {title} {High-entropy materials for
  catalysis: A new frontier},}\ }\href
  {https://www.science.org/doi/full/10.1126/sciadv.abg1600} {\bibfield
  {journal} {\bibinfo  {journal} {Science Advances}\ }\textbf {\bibinfo
  {volume} {7}},\ \bibinfo {pages} {eabg1600} (\bibinfo {year}
  {2021})}\BibitemShut {NoStop}%
\bibitem [{\citenamefont {Sickafus}\ \emph {et~al.}(1999)\citenamefont
  {Sickafus}, \citenamefont {Wills},\ and\ \citenamefont
  {Grimes}}]{sickafus1999structure}%
  \BibitemOpen
  \bibfield  {author} {\bibinfo {author} {\bibfnamefont {Kurt~E}\ \bibnamefont
  {Sickafus}}, \bibinfo {author} {\bibfnamefont {John~M}\ \bibnamefont
  {Wills}}, \ and\ \bibinfo {author} {\bibfnamefont {Norman~W}\ \bibnamefont
  {Grimes}},\ }\bibfield  {title} {\enquote {\bibinfo {title} {Structure of
  spinel},}\ }\href {https://doi.org/10.1111/j.1151-2916.1999.tb02241.x}
  {\bibfield  {journal} {\bibinfo  {journal} {Journal of the American Ceramic
  Society}\ }\textbf {\bibinfo {volume} {82}},\ \bibinfo {pages} {3279--3292}
  (\bibinfo {year} {1999})}\BibitemShut {NoStop}%
\bibitem [{\citenamefont {Zhao}\ \emph {et~al.}(2017)\citenamefont {Zhao},
  \citenamefont {Yan}, \citenamefont {Chen},\ and\ \citenamefont
  {Chen}}]{zhao2017spinels}%
  \BibitemOpen
  \bibfield  {author} {\bibinfo {author} {\bibfnamefont {Qing}\ \bibnamefont
  {Zhao}}, \bibinfo {author} {\bibfnamefont {Zhenhua}\ \bibnamefont {Yan}},
  \bibinfo {author} {\bibfnamefont {Chengcheng}\ \bibnamefont {Chen}}, \ and\
  \bibinfo {author} {\bibfnamefont {Jun}\ \bibnamefont {Chen}},\ }\bibfield
  {title} {\enquote {\bibinfo {title} {Spinels: controlled preparation, oxygen
  reduction/evolution reaction application, and beyond},}\ }\href
  {https://pubs.acs.org/doi/10.1021/acs.chemrev.7b00051} {\bibfield  {journal}
  {\bibinfo  {journal} {Chemical reviews}\ }\textbf {\bibinfo {volume} {117}},\
  \bibinfo {pages} {10121--10211} (\bibinfo {year} {2017})}\BibitemShut
  {NoStop}%
\bibitem [{\citenamefont {Verwey}\ and\ \citenamefont
  {Heilmann}(1947)}]{verwey1947physical}%
  \BibitemOpen
  \bibfield  {author} {\bibinfo {author} {\bibfnamefont {EJW}\ \bibnamefont
  {Verwey}}\ and\ \bibinfo {author} {\bibfnamefont {EL}~\bibnamefont
  {Heilmann}},\ }\bibfield  {title} {\enquote {\bibinfo {title} {Physical
  properties and cation arrangement of oxides with spinel structures i. cation
  arrangement in spinels},}\ }\href
  {https://aip.scitation.org/doi/abs/10.1063/1.1746464} {\bibfield  {journal}
  {\bibinfo  {journal} {The Journal of Chemical Physics}\ }\textbf {\bibinfo
  {volume} {15}},\ \bibinfo {pages} {174--180} (\bibinfo {year}
  {1947})}\BibitemShut {NoStop}%
\bibitem [{\citenamefont {O'Neill}\ and\ \citenamefont
  {Navrotsky}(1983)}]{o1983simple}%
  \BibitemOpen
  \bibfield  {author} {\bibinfo {author} {\bibfnamefont {Hugh St~C}\
  \bibnamefont {O'Neill}}\ and\ \bibinfo {author} {\bibfnamefont {Alexandra}\
  \bibnamefont {Navrotsky}},\ }\bibfield  {title} {\enquote {\bibinfo {title}
  {Simple spinels: crystallographic parameters, cation radii, lattice energies,
  and cation distribution},}\ }\href
  {https://pubs.geoscienceworld.org/msa/ammin/article-pdf/68/1-2/181/4208815/am68_181.pdf}
  {\bibfield  {journal} {\bibinfo  {journal} {American Mineralogist}\ }\textbf
  {\bibinfo {volume} {68}},\ \bibinfo {pages} {181--194} (\bibinfo {year}
  {1983})}\BibitemShut {NoStop}%
\bibitem [{\citenamefont {Song}\ \emph {et~al.}(2012)\citenamefont {Song},
  \citenamefont {Shin}, \citenamefont {Lu}, \citenamefont {Amos}, \citenamefont
  {Manthiram},\ and\ \citenamefont {Goodenough}}]{song2012role}%
  \BibitemOpen
  \bibfield  {author} {\bibinfo {author} {\bibfnamefont {Jie}\ \bibnamefont
  {Song}}, \bibinfo {author} {\bibfnamefont {Dong~Wook}\ \bibnamefont {Shin}},
  \bibinfo {author} {\bibfnamefont {Yuhao}\ \bibnamefont {Lu}}, \bibinfo
  {author} {\bibfnamefont {Charles~D}\ \bibnamefont {Amos}}, \bibinfo {author}
  {\bibfnamefont {Arumugam}\ \bibnamefont {Manthiram}}, \ and\ \bibinfo
  {author} {\bibfnamefont {John~B}\ \bibnamefont {Goodenough}},\ }\bibfield
  {title} {\enquote {\bibinfo {title} {Role of oxygen vacancies on the
  performance of {Li[Ni$_{0.5-x}$Mn$_{1.5+x}$] O$_4$ ($x= 0$, 0.05, and 0.08)}
  spinel cathodes for lithium-ion batteries},}\ }\href
  {https://doi.org/10.1021/cm301825h} {\bibfield  {journal} {\bibinfo
  {journal} {Chemistry of Materials}\ }\textbf {\bibinfo {volume} {24}},\
  \bibinfo {pages} {3101--3109} (\bibinfo {year} {2012})}\BibitemShut {NoStop}%
\bibitem [{\citenamefont {Torres}\ \emph {et~al.}(2014)\citenamefont {Torres},
  \citenamefont {Pasquevich}, \citenamefont {Z{\'e}lis}, \citenamefont
  {Golmar}, \citenamefont {Heluani}, \citenamefont {Nayak}, \citenamefont
  {Adeagbo}, \citenamefont {Hergert}, \citenamefont {Hoffmann}, \citenamefont
  {Ernst} \emph {et~al.}}]{torres2014oxygen}%
  \BibitemOpen
  \bibfield  {author} {\bibinfo {author} {\bibfnamefont {CE~Rodr{\'\i}guez}\
  \bibnamefont {Torres}}, \bibinfo {author} {\bibfnamefont {Gustavo~Alberto}\
  \bibnamefont {Pasquevich}}, \bibinfo {author} {\bibfnamefont {P~Mendoza}\
  \bibnamefont {Z{\'e}lis}}, \bibinfo {author} {\bibfnamefont {Federico}\
  \bibnamefont {Golmar}}, \bibinfo {author} {\bibfnamefont {SP}~\bibnamefont
  {Heluani}}, \bibinfo {author} {\bibfnamefont {Sanjeev~K}\ \bibnamefont
  {Nayak}}, \bibinfo {author} {\bibfnamefont {Waheed~A}\ \bibnamefont
  {Adeagbo}}, \bibinfo {author} {\bibfnamefont {Wolfram}\ \bibnamefont
  {Hergert}}, \bibinfo {author} {\bibfnamefont {Martin}\ \bibnamefont
  {Hoffmann}}, \bibinfo {author} {\bibfnamefont {Arthur}\ \bibnamefont
  {Ernst}},  \emph {et~al.},\ }\bibfield  {title} {\enquote {\bibinfo {title}
  {Oxygen-vacancy-induced local ferromagnetism as a driving mechanism in
  enhancing the magnetic response of ferrites},}\ }\href
  {https://journals-aps-org.eu1.proxy.openathens.net/prb/abstract/10.1103/PhysRevB.89.104411}
  {\bibfield  {journal} {\bibinfo  {journal} {Physical Review B}\ }\textbf
  {\bibinfo {volume} {89}},\ \bibinfo {pages} {104411} (\bibinfo {year}
  {2014})}\BibitemShut {NoStop}%
\bibitem [{\citenamefont {O’Quinn}\ \emph {et~al.}(2017)\citenamefont
  {O’Quinn}, \citenamefont {Shamblin}, \citenamefont {Perlov}, \citenamefont
  {Ewing}, \citenamefont {Neuefeind}, \citenamefont {Feygenson}, \citenamefont
  {Gussev},\ and\ \citenamefont {Lang}}]{o2017inversion}%
  \BibitemOpen
  \bibfield  {author} {\bibinfo {author} {\bibfnamefont {Eric~C}\ \bibnamefont
  {O’Quinn}}, \bibinfo {author} {\bibfnamefont {Jacob}\ \bibnamefont
  {Shamblin}}, \bibinfo {author} {\bibfnamefont {Brandon}\ \bibnamefont
  {Perlov}}, \bibinfo {author} {\bibfnamefont {Rodney~C}\ \bibnamefont
  {Ewing}}, \bibinfo {author} {\bibfnamefont {Joerg}\ \bibnamefont
  {Neuefeind}}, \bibinfo {author} {\bibfnamefont {Mikhail}\ \bibnamefont
  {Feygenson}}, \bibinfo {author} {\bibfnamefont {Igor}\ \bibnamefont
  {Gussev}}, \ and\ \bibinfo {author} {\bibfnamefont {Maik}\ \bibnamefont
  {Lang}},\ }\bibfield  {title} {\enquote {\bibinfo {title} {Inversion in
  {Mg$_{1-x}$Ni$_x$Al$_2$O$_4$} spinel: New insight into local structure},}\
  }\href {https://pubs.acs.org/doi/abs/10.1021/jacs.7b04370} {\bibfield
  {journal} {\bibinfo  {journal} {Journal of the American Chemical Society}\
  }\textbf {\bibinfo {volume} {139}},\ \bibinfo {pages} {10395--10402}
  (\bibinfo {year} {2017})}\BibitemShut {NoStop}%
\bibitem [{\citenamefont {Liu}\ \emph {et~al.}(2019)\citenamefont {Liu},
  \citenamefont {Wang}, \citenamefont {Borkiewicz}, \citenamefont {Hu},
  \citenamefont {Xiao}, \citenamefont {Chen},\ and\ \citenamefont
  {Page}}]{liu2019unified}%
  \BibitemOpen
  \bibfield  {author} {\bibinfo {author} {\bibfnamefont {Jue}\ \bibnamefont
  {Liu}}, \bibinfo {author} {\bibfnamefont {Xuelong}\ \bibnamefont {Wang}},
  \bibinfo {author} {\bibfnamefont {Olaf~J}\ \bibnamefont {Borkiewicz}},
  \bibinfo {author} {\bibfnamefont {Enyuan}\ \bibnamefont {Hu}}, \bibinfo
  {author} {\bibfnamefont {Rui-Juan}\ \bibnamefont {Xiao}}, \bibinfo {author}
  {\bibfnamefont {Liquan}\ \bibnamefont {Chen}}, \ and\ \bibinfo {author}
  {\bibfnamefont {Katharine}\ \bibnamefont {Page}},\ }\bibfield  {title}
  {\enquote {\bibinfo {title} {Unified view of the local cation-ordered state
  in inverse spinel oxides},}\ }\href
  {https://doi.org/10.1021/acs.inorgchem.9b01685} {\bibfield  {journal}
  {\bibinfo  {journal} {Inorganic Chemistry}\ }\textbf {\bibinfo {volume}
  {58}},\ \bibinfo {pages} {14389--14402} (\bibinfo {year} {2019})}\BibitemShut
  {NoStop}%
\bibitem [{\citenamefont {Navrotsky}\ and\ \citenamefont
  {Kleppa}(1967)}]{navrotsky1967thermodynamics}%
  \BibitemOpen
  \bibfield  {author} {\bibinfo {author} {\bibfnamefont {Alexandra}\
  \bibnamefont {Navrotsky}}\ and\ \bibinfo {author} {\bibfnamefont
  {OJ}~\bibnamefont {Kleppa}},\ }\bibfield  {title} {\enquote {\bibinfo {title}
  {The thermodynamics of cation distributions in simple spinels},}\ }\href
  {https://www.sciencedirect.com/science/article/pii/0022190267800083}
  {\bibfield  {journal} {\bibinfo  {journal} {Journal of Inorganic and nuclear
  Chemistry}\ }\textbf {\bibinfo {volume} {29}},\ \bibinfo {pages} {2701--2714}
  (\bibinfo {year} {1967})}\BibitemShut {NoStop}%
\bibitem [{\citenamefont {Navrotsky}\ and\ \citenamefont
  {Kleppa}(1968)}]{navrotsky1968thermodynamics}%
  \BibitemOpen
  \bibfield  {author} {\bibinfo {author} {\bibfnamefont {A}~\bibnamefont
  {Navrotsky}}\ and\ \bibinfo {author} {\bibfnamefont {OJ}~\bibnamefont
  {Kleppa}},\ }\bibfield  {title} {\enquote {\bibinfo {title} {Thermodynamics
  of formation of simple spinels},}\ }\href
  {https://www.sciencedirect.com/science/article/pii/0022190268804750}
  {\bibfield  {journal} {\bibinfo  {journal} {Journal of Inorganic and Nuclear
  Chemistry}\ }\textbf {\bibinfo {volume} {30}},\ \bibinfo {pages} {479--498}
  (\bibinfo {year} {1968})}\BibitemShut {NoStop}%
\bibitem [{\citenamefont {Navrotsky}(1969)}]{navrotsky1969thermodynamics}%
  \BibitemOpen
  \bibfield  {author} {\bibinfo {author} {\bibfnamefont {A}~\bibnamefont
  {Navrotsky}},\ }\bibfield  {title} {\enquote {\bibinfo {title}
  {Thermodynamics of {$A_3$O$_4$-$B_3$O$_4$} spinel solid solutions},}\ }\href
  {https://doi.org/10.1016/0022-1902(69)80054-0} {\bibfield  {journal}
  {\bibinfo  {journal} {Journal of Inorganic and Nuclear Chemistry}\ }\textbf
  {\bibinfo {volume} {31}},\ \bibinfo {pages} {59--72} (\bibinfo {year}
  {1969})}\BibitemShut {NoStop}%
\bibitem [{\citenamefont {Zhang}\ \emph {et~al.}(2019)\citenamefont {Zhang},
  \citenamefont {Yan}, \citenamefont {Calder}, \citenamefont {Zheng},
  \citenamefont {McGuire}, \citenamefont {Abernathy}, \citenamefont {Ren},
  \citenamefont {Lapidus}, \citenamefont {Page}, \citenamefont {Zheng},
  \citenamefont {Freeland}, \citenamefont {Budai},\ and\ \citenamefont
  {Hermann}}]{ChemOfMat-Rocksalt-magnetism}%
  \BibitemOpen
  \bibfield  {author} {\bibinfo {author} {\bibfnamefont {Junjie}\ \bibnamefont
  {Zhang}}, \bibinfo {author} {\bibfnamefont {Jiaqiang}\ \bibnamefont {Yan}},
  \bibinfo {author} {\bibfnamefont {Stuart}\ \bibnamefont {Calder}}, \bibinfo
  {author} {\bibfnamefont {Qiang}\ \bibnamefont {Zheng}}, \bibinfo {author}
  {\bibfnamefont {Michael~A.}\ \bibnamefont {McGuire}}, \bibinfo {author}
  {\bibfnamefont {Douglas~L.}\ \bibnamefont {Abernathy}}, \bibinfo {author}
  {\bibfnamefont {Yang}\ \bibnamefont {Ren}}, \bibinfo {author} {\bibfnamefont
  {Saul~H.}\ \bibnamefont {Lapidus}}, \bibinfo {author} {\bibfnamefont
  {Katharine}\ \bibnamefont {Page}}, \bibinfo {author} {\bibfnamefont {Hong}\
  \bibnamefont {Zheng}}, \bibinfo {author} {\bibfnamefont {John~W.}\
  \bibnamefont {Freeland}}, \bibinfo {author} {\bibfnamefont {John~D.}\
  \bibnamefont {Budai}}, \ and\ \bibinfo {author} {\bibfnamefont {Raphael~P.}\
  \bibnamefont {Hermann}},\ }\bibfield  {title} {\enquote {\bibinfo {title}
  {Long-range antiferromagnetic order in a rocksalt high entropy oxide},}\
  }\href {\doibase 10.1021/acs.chemmater.9b00624} {\bibfield  {journal}
  {\bibinfo  {journal} {Chemistry of Materials}\ }\textbf {\bibinfo {volume}
  {31}},\ \bibinfo {pages} {3705--3711} (\bibinfo {year} {2019})}\BibitemShut
  {NoStop}%
\bibitem [{\citenamefont {Dąbrowa}\ \emph {et~al.}(2018)\citenamefont
  {Dąbrowa}, \citenamefont {Stygar}, \citenamefont {Mikuła}, \citenamefont
  {Knapik}, \citenamefont {Mroczka}, \citenamefont {Tejchman}, \citenamefont
  {Danielewski},\ and\ \citenamefont {Martin}}]{s-HEO-first-report}%
  \BibitemOpen
  \bibfield  {author} {\bibinfo {author} {\bibfnamefont {Juliusz}\ \bibnamefont
  {Dąbrowa}}, \bibinfo {author} {\bibfnamefont {Mirosław}\ \bibnamefont
  {Stygar}}, \bibinfo {author} {\bibfnamefont {Andrzej}\ \bibnamefont
  {Mikuła}}, \bibinfo {author} {\bibfnamefont {Arkadiusz}\ \bibnamefont
  {Knapik}}, \bibinfo {author} {\bibfnamefont {Krzysztof}\ \bibnamefont
  {Mroczka}}, \bibinfo {author} {\bibfnamefont {Waldemar}\ \bibnamefont
  {Tejchman}}, \bibinfo {author} {\bibfnamefont {Marek}\ \bibnamefont
  {Danielewski}}, \ and\ \bibinfo {author} {\bibfnamefont {Manfred}\
  \bibnamefont {Martin}},\ }\bibfield  {title} {\enquote {\bibinfo {title}
  {Synthesis and microstructure of the (co,cr,fe,mn,ni)3o4 high entropy oxide
  characterized by spinel structure},}\ }\href {\doibase
  https://doi.org/10.1016/j.matlet.2017.12.148} {\bibfield  {journal} {\bibinfo
   {journal} {Materials Letters}\ }\textbf {\bibinfo {volume} {216}},\ \bibinfo
  {pages} {32--36} (\bibinfo {year} {2018})}\BibitemShut {NoStop}%
\bibitem [{\citenamefont {Cieslak}\ \emph
  {et~al.}(2021{\natexlab{a}})\citenamefont {Cieslak}, \citenamefont
  {Reissner}, \citenamefont {Berent}, \citenamefont {Dabrowa}, \citenamefont
  {Stygar}, \citenamefont {Mozdzierz},\ and\ \citenamefont
  {Zajusz}}]{Spinel-HEO-Cieslak-FiM-Mossbauer}%
  \BibitemOpen
  \bibfield  {author} {\bibinfo {author} {\bibfnamefont {J.}~\bibnamefont
  {Cieslak}}, \bibinfo {author} {\bibfnamefont {M.}~\bibnamefont {Reissner}},
  \bibinfo {author} {\bibfnamefont {K.}~\bibnamefont {Berent}}, \bibinfo
  {author} {\bibfnamefont {J.}~\bibnamefont {Dabrowa}}, \bibinfo {author}
  {\bibfnamefont {M.}~\bibnamefont {Stygar}}, \bibinfo {author} {\bibfnamefont
  {M.}~\bibnamefont {Mozdzierz}}, \ and\ \bibinfo {author} {\bibfnamefont
  {M.}~\bibnamefont {Zajusz}},\ }\bibfield  {title} {\enquote {\bibinfo {title}
  {Magnetic properties and ionic distribution in high entropy spinels studied
  by mössbauer and ab initio methods},}\ }\href {\doibase
  https://doi.org/10.1016/j.actamat.2020.116600} {\bibfield  {journal}
  {\bibinfo  {journal} {Acta Materialia}\ }\textbf {\bibinfo {volume} {206}},\
  \bibinfo {pages} {116600} (\bibinfo {year} {2021}{\natexlab{a}})}\BibitemShut
  {NoStop}%
\bibitem [{\citenamefont {Mao}\ \emph {et~al.}(2019)\citenamefont {Mao},
  \citenamefont {Quan}, \citenamefont {Xiang}, \citenamefont {Zhang},
  \citenamefont {Kuramoto},\ and\ \citenamefont {Xia}}]{Spinel-HEO-Mao-FiM}%
  \BibitemOpen
  \bibfield  {author} {\bibinfo {author} {\bibfnamefont {Aiqin}\ \bibnamefont
  {Mao}}, \bibinfo {author} {\bibfnamefont {Feng}\ \bibnamefont {Quan}},
  \bibinfo {author} {\bibfnamefont {Hou-Zheng}\ \bibnamefont {Xiang}}, \bibinfo
  {author} {\bibfnamefont {Zhan-Guo}\ \bibnamefont {Zhang}}, \bibinfo {author}
  {\bibfnamefont {Koji}\ \bibnamefont {Kuramoto}}, \ and\ \bibinfo {author}
  {\bibfnamefont {Ai-Lin}\ \bibnamefont {Xia}},\ }\bibfield  {title} {\enquote
  {\bibinfo {title} {Facile synthesis and ferrimagnetic property of spinel
  {(CoCrFeMnNi)$_3$O$_4$} high-entropy oxide nanocrystalline powder},}\ }\href
  {\doibase https://doi.org/10.1016/j.molstruc.2019.05.073} {\bibfield
  {journal} {\bibinfo  {journal} {Journal of Molecular Structure}\ }\textbf
  {\bibinfo {volume} {1194}},\ \bibinfo {pages} {11--18} (\bibinfo {year}
  {2019})}\BibitemShut {NoStop}%
\bibitem [{\citenamefont {Grzesik}\ \emph {et~al.}(2020)\citenamefont
  {Grzesik}, \citenamefont {Smoła}, \citenamefont {Miszczak}, \citenamefont
  {Stygar}, \citenamefont {Dąbrowa}, \citenamefont {Zajusz}, \citenamefont
  {Świerczek},\ and\ \citenamefont
  {Danielewski}}]{Spinel-HEO-Grzesik-diffusion}%
  \BibitemOpen
  \bibfield  {author} {\bibinfo {author} {\bibfnamefont {Zbigniew}\
  \bibnamefont {Grzesik}}, \bibinfo {author} {\bibfnamefont {Grzegorz}\
  \bibnamefont {Smoła}}, \bibinfo {author} {\bibfnamefont {Maria}\
  \bibnamefont {Miszczak}}, \bibinfo {author} {\bibfnamefont {Mirosław}\
  \bibnamefont {Stygar}}, \bibinfo {author} {\bibfnamefont {Juliusz}\
  \bibnamefont {Dąbrowa}}, \bibinfo {author} {\bibfnamefont {Marek}\
  \bibnamefont {Zajusz}}, \bibinfo {author} {\bibfnamefont {Konrad}\
  \bibnamefont {Świerczek}}, \ and\ \bibinfo {author} {\bibfnamefont {Marek}\
  \bibnamefont {Danielewski}},\ }\bibfield  {title} {\enquote {\bibinfo {title}
  {Defect structure and transport properties of {(Co,Cr,Fe,Mn,Ni)$_3$O$_4$}
  spinel-structured high entropy oxide},}\ }\href {\doibase
  https://doi.org/10.1016/j.jeurceramsoc.2019.10.026} {\bibfield  {journal}
  {\bibinfo  {journal} {Journal of the European Ceramic Society}\ }\textbf
  {\bibinfo {volume} {40}},\ \bibinfo {pages} {835--839} (\bibinfo {year}
  {2020})}\BibitemShut {NoStop}%
\bibitem [{\citenamefont {Huang}\ \emph {et~al.}(2021)\citenamefont {Huang},
  \citenamefont {Huang}, \citenamefont {Wu}, \citenamefont {Patra},
  \citenamefont {{Xuyen Nguyen}}, \citenamefont {Chang}, \citenamefont
  {Clemens}, \citenamefont {Ting}, \citenamefont {Li}, \citenamefont {Chang},\
  and\ \citenamefont {Wu}}]{Spinel-HEO-Huang-Lithiation}%
  \BibitemOpen
  \bibfield  {author} {\bibinfo {author} {\bibfnamefont {Chih-Yang}\
  \bibnamefont {Huang}}, \bibinfo {author} {\bibfnamefont {Chun-Wei}\
  \bibnamefont {Huang}}, \bibinfo {author} {\bibfnamefont {Min-Ci}\
  \bibnamefont {Wu}}, \bibinfo {author} {\bibfnamefont {Jagabandhu}\
  \bibnamefont {Patra}}, \bibinfo {author} {\bibfnamefont {Thi}\ \bibnamefont
  {{Xuyen Nguyen}}}, \bibinfo {author} {\bibfnamefont {Mu-Tung}\ \bibnamefont
  {Chang}}, \bibinfo {author} {\bibfnamefont {Oliver}\ \bibnamefont {Clemens}},
  \bibinfo {author} {\bibfnamefont {Jyh-Ming}\ \bibnamefont {Ting}}, \bibinfo
  {author} {\bibfnamefont {Ju}~\bibnamefont {Li}}, \bibinfo {author}
  {\bibfnamefont {Jeng-Kuei}\ \bibnamefont {Chang}}, \ and\ \bibinfo {author}
  {\bibfnamefont {Wen-Wei}\ \bibnamefont {Wu}},\ }\bibfield  {title} {\enquote
  {\bibinfo {title} {Atomic-scale investigation of lithiation/delithiation
  mechanism in high-entropy spinel oxide with superior electrochemical
  performance},}\ }\href {\doibase https://doi.org/10.1016/j.cej.2021.129838}
  {\bibfield  {journal} {\bibinfo  {journal} {Chemical Engineering Journal}\
  }\textbf {\bibinfo {volume} {420}},\ \bibinfo {pages} {129838} (\bibinfo
  {year} {2021})}\BibitemShut {NoStop}%
\bibitem [{\citenamefont {Stygar}\ \emph {et~al.}(2020)\citenamefont {Stygar},
  \citenamefont {Dąbrowa}, \citenamefont {Moździerz}, \citenamefont {Zajusz},
  \citenamefont {Skubida}, \citenamefont {Mroczka}, \citenamefont {Berent},
  \citenamefont {Świerczek},\ and\ \citenamefont
  {Danielewski}}]{Spinel-HEO-Hygar-Seebeck}%
  \BibitemOpen
  \bibfield  {author} {\bibinfo {author} {\bibfnamefont {Mirosław}\
  \bibnamefont {Stygar}}, \bibinfo {author} {\bibfnamefont {Juliusz}\
  \bibnamefont {Dąbrowa}}, \bibinfo {author} {\bibfnamefont {Maciej}\
  \bibnamefont {Moździerz}}, \bibinfo {author} {\bibfnamefont {Marek}\
  \bibnamefont {Zajusz}}, \bibinfo {author} {\bibfnamefont {Wojciech}\
  \bibnamefont {Skubida}}, \bibinfo {author} {\bibfnamefont {Krzysztof}\
  \bibnamefont {Mroczka}}, \bibinfo {author} {\bibfnamefont {Katarzyna}\
  \bibnamefont {Berent}}, \bibinfo {author} {\bibfnamefont {Konrad}\
  \bibnamefont {Świerczek}}, \ and\ \bibinfo {author} {\bibfnamefont {Marek}\
  \bibnamefont {Danielewski}},\ }\bibfield  {title} {\enquote {\bibinfo {title}
  {Formation and properties of high entropy oxides in {Co-Cr-Fe-Mg-Mn-Ni-O}
  system: Novel {(Cr,Fe,Mg,Mn,Ni)$_3$O$_4$ and (Co,Cr,Fe,Mg,Mn)$_3$O$_4$} high
  entropy spinels},}\ }\href {\doibase
  https://doi.org/10.1016/j.jeurceramsoc.2019.11.030} {\bibfield  {journal}
  {\bibinfo  {journal} {Journal of the European Ceramic Society}\ }\textbf
  {\bibinfo {volume} {40}},\ \bibinfo {pages} {1644--1650} (\bibinfo {year}
  {2020})}\BibitemShut {NoStop}%
\bibitem [{\citenamefont {Nguyen}\ \emph {et~al.}(2020)\citenamefont {Nguyen},
  \citenamefont {Patra}, \citenamefont {Chang},\ and\ \citenamefont
  {Ting}}]{Spinel-HEO-Nguyen-batteries}%
  \BibitemOpen
  \bibfield  {author} {\bibinfo {author} {\bibfnamefont {Thi~Xuyen}\
  \bibnamefont {Nguyen}}, \bibinfo {author} {\bibfnamefont {Jagabandhu}\
  \bibnamefont {Patra}}, \bibinfo {author} {\bibfnamefont {Jeng-Kuei}\
  \bibnamefont {Chang}}, \ and\ \bibinfo {author} {\bibfnamefont {Jyh-Ming}\
  \bibnamefont {Ting}},\ }\bibfield  {title} {\enquote {\bibinfo {title} {High
  entropy spinel oxide nanoparticles for superior lithiation–delithiation
  performance},}\ }\href {\doibase 10.1039/D0TA04844E} {\bibfield  {journal}
  {\bibinfo  {journal} {J. Mater. Chem. A}\ }\textbf {\bibinfo {volume} {8}},\
  \bibinfo {pages} {18963--18973} (\bibinfo {year} {2020})}\BibitemShut
  {NoStop}%
\bibitem [{\citenamefont {Talluri}\ \emph {et~al.}(2021)\citenamefont
  {Talluri}, \citenamefont {Aparna}, \citenamefont {Sreenivasulu},
  \citenamefont {Bhattacharya},\ and\ \citenamefont
  {Thomas}}]{Spinel-HEO-Talluri-supercapacitor}%
  \BibitemOpen
  \bibfield  {author} {\bibinfo {author} {\bibfnamefont {Bhusankar}\
  \bibnamefont {Talluri}}, \bibinfo {author} {\bibfnamefont {M.L.}\
  \bibnamefont {Aparna}}, \bibinfo {author} {\bibfnamefont {N.}~\bibnamefont
  {Sreenivasulu}}, \bibinfo {author} {\bibfnamefont {S.S.}\ \bibnamefont
  {Bhattacharya}}, \ and\ \bibinfo {author} {\bibfnamefont {Tiju}\ \bibnamefont
  {Thomas}},\ }\bibfield  {title} {\enquote {\bibinfo {title} {High entropy
  spinel metal oxide {(CoCrFeMnNi)$_3$O$_4$} nanoparticles as a
  high-performance supercapacitor electrode material},}\ }\href {\doibase
  https://doi.org/10.1016/j.est.2021.103004} {\bibfield  {journal} {\bibinfo
  {journal} {Journal of Energy Storage}\ }\textbf {\bibinfo {volume} {42}},\
  \bibinfo {pages} {103004} (\bibinfo {year} {2021})}\BibitemShut {NoStop}%
\bibitem [{\citenamefont {Wang}\ \emph {et~al.}(2020)\citenamefont {Wang},
  \citenamefont {Jiang}, \citenamefont {Duan}, \citenamefont {Mao},
  \citenamefont {Dong}, \citenamefont {Dong}, \citenamefont {Wang},
  \citenamefont {Luo}, \citenamefont {Liu},\ and\ \citenamefont
  {Qi}}]{Spinel-HEO-Wang-batteries}%
  \BibitemOpen
  \bibfield  {author} {\bibinfo {author} {\bibfnamefont {Dan}\ \bibnamefont
  {Wang}}, \bibinfo {author} {\bibfnamefont {Shunda}\ \bibnamefont {Jiang}},
  \bibinfo {author} {\bibfnamefont {Chanqin}\ \bibnamefont {Duan}}, \bibinfo
  {author} {\bibfnamefont {Jing}\ \bibnamefont {Mao}}, \bibinfo {author}
  {\bibfnamefont {Ying}\ \bibnamefont {Dong}}, \bibinfo {author} {\bibfnamefont
  {Kangze}\ \bibnamefont {Dong}}, \bibinfo {author} {\bibfnamefont {Zhiyuan}\
  \bibnamefont {Wang}}, \bibinfo {author} {\bibfnamefont {Shaohua}\
  \bibnamefont {Luo}}, \bibinfo {author} {\bibfnamefont {Yanguo}\ \bibnamefont
  {Liu}}, \ and\ \bibinfo {author} {\bibfnamefont {Xiwei}\ \bibnamefont {Qi}},\
  }\bibfield  {title} {\enquote {\bibinfo {title} {Spinel-structured high
  entropy oxide {(FeCoNiCrMn)$_3$O$_4$} as anode towards superior lithium
  storage performance},}\ }\href {\doibase
  https://doi.org/10.1016/j.jallcom.2020.156158} {\bibfield  {journal}
  {\bibinfo  {journal} {Journal of Alloys and Compounds}\ }\textbf {\bibinfo
  {volume} {844}},\ \bibinfo {pages} {156158} (\bibinfo {year}
  {2020})}\BibitemShut {NoStop}%
\bibitem [{\citenamefont {Wang}\ \emph {et~al.}(2019)\citenamefont {Wang},
  \citenamefont {Liu}, \citenamefont {Du}, \citenamefont {Zhang}, \citenamefont
  {Li}, \citenamefont {Xiao}, \citenamefont {Chen}, \citenamefont {Chen},
  \citenamefont {Wang}, \citenamefont {Zou},\ and\ \citenamefont
  {Wang}}]{Spinel-HEO-Wang-water}%
  \BibitemOpen
  \bibfield  {author} {\bibinfo {author} {\bibfnamefont {Dongdong}\
  \bibnamefont {Wang}}, \bibinfo {author} {\bibfnamefont {Zhijuan}\
  \bibnamefont {Liu}}, \bibinfo {author} {\bibfnamefont {Shiqian}\ \bibnamefont
  {Du}}, \bibinfo {author} {\bibfnamefont {Yiqiong}\ \bibnamefont {Zhang}},
  \bibinfo {author} {\bibfnamefont {Hao}\ \bibnamefont {Li}}, \bibinfo {author}
  {\bibfnamefont {Zhaohui}\ \bibnamefont {Xiao}}, \bibinfo {author}
  {\bibfnamefont {Wei}\ \bibnamefont {Chen}}, \bibinfo {author} {\bibfnamefont
  {Ru}~\bibnamefont {Chen}}, \bibinfo {author} {\bibfnamefont {Yanyong}\
  \bibnamefont {Wang}}, \bibinfo {author} {\bibfnamefont {Yuqin}\ \bibnamefont
  {Zou}}, \ and\ \bibinfo {author} {\bibfnamefont {Shuangyin}\ \bibnamefont
  {Wang}},\ }\bibfield  {title} {\enquote {\bibinfo {title} {Low-temperature
  synthesis of small-sized high-entropy oxides for water oxidation},}\ }\href
  {\doibase 10.1039/C9TA08740K} {\bibfield  {journal} {\bibinfo  {journal} {J.
  Mater. Chem. A}\ }\textbf {\bibinfo {volume} {7}},\ \bibinfo {pages}
  {24211--24216} (\bibinfo {year} {2019})}\BibitemShut {NoStop}%
\bibitem [{\citenamefont {Sarkar}\ \emph {et~al.}(2021)\citenamefont {Sarkar},
  \citenamefont {Eggert}, \citenamefont {Witte}, \citenamefont {Lill},
  \citenamefont {Velasco}, \citenamefont {Wang}, \citenamefont {Sonar},
  \citenamefont {Ollefs}, \citenamefont {Bhattacharya}, \citenamefont {Brand}
  \emph {et~al.}}]{sarkar2021comprehensive}%
  \BibitemOpen
  \bibfield  {author} {\bibinfo {author} {\bibfnamefont {Abhishek}\
  \bibnamefont {Sarkar}}, \bibinfo {author} {\bibfnamefont {Benedikt}\
  \bibnamefont {Eggert}}, \bibinfo {author} {\bibfnamefont {Ralf}\ \bibnamefont
  {Witte}}, \bibinfo {author} {\bibfnamefont {Johanna}\ \bibnamefont {Lill}},
  \bibinfo {author} {\bibfnamefont {Leonardo}\ \bibnamefont {Velasco}},
  \bibinfo {author} {\bibfnamefont {Qingsong}\ \bibnamefont {Wang}}, \bibinfo
  {author} {\bibfnamefont {Janhavika}\ \bibnamefont {Sonar}}, \bibinfo {author}
  {\bibfnamefont {Katharina}\ \bibnamefont {Ollefs}}, \bibinfo {author}
  {\bibfnamefont {Subramshu~S}\ \bibnamefont {Bhattacharya}}, \bibinfo {author}
  {\bibfnamefont {Richard~A}\ \bibnamefont {Brand}},  \emph {et~al.},\
  }\bibfield  {title} {\enquote {\bibinfo {title} {Comprehensive investigation
  of crystallographic, spin-electronic and magnetic structure of
  {(Co$_{0.2}$Cr$_{0.2}$Fe$_{0.2}$Mn$_{0.2}$Ni$_{0.2}$)$_3$O$_4$}: Unraveling
  the suppression of configuration entropy in high entropy oxides},}\ }\href
  {https://www.sciencedirect.com/science/article/pii/S1359645421009599}
  {\bibfield  {journal} {\bibinfo  {journal} {Acta Materialia}\ ,\ \bibinfo
  {pages} {117581}} (\bibinfo {year} {2021})}\BibitemShut {NoStop}%
\bibitem [{\citenamefont {Casado}\ and\ \citenamefont
  {Rasines}(1982)}]{Casado-MnGa2O4}%
  \BibitemOpen
  \bibfield  {author} {\bibinfo {author} {\bibfnamefont {P~Garcia}\
  \bibnamefont {Casado}}\ and\ \bibinfo {author} {\bibfnamefont
  {I}~\bibnamefont {Rasines}},\ }\bibfield  {title} {\enquote {\bibinfo {title}
  {Crystal data for the spinels {$M$Ga$_2$O$_4$ ($M=$~Mg, Mn)}},}\ }\href@noop
  {} {\bibfield  {journal} {\bibinfo  {journal} {Zeitschrift f{\"u}r
  Kristallographie}\ }\textbf {\bibinfo {volume} {160}},\ \bibinfo {pages}
  {33--37} (\bibinfo {year} {1982})}\BibitemShut {NoStop}%
\bibitem [{\citenamefont {Ghose}\ \emph {et~al.}(1977)\citenamefont {Ghose},
  \citenamefont {Hallam},\ and\ \citenamefont {Read}}]{Ghose-FeGa2O4}%
  \BibitemOpen
  \bibfield  {author} {\bibinfo {author} {\bibfnamefont {J}~\bibnamefont
  {Ghose}}, \bibinfo {author} {\bibfnamefont {G~C}\ \bibnamefont {Hallam}}, \
  and\ \bibinfo {author} {\bibfnamefont {D~A}\ \bibnamefont {Read}},\
  }\bibfield  {title} {\enquote {\bibinfo {title} {A magnetic study of
  {FeGa$_2$O$_4$}},}\ }\href {\doibase 10.1088/0022-3719/10/7/014} {\bibfield
  {journal} {\bibinfo  {journal} {Journal of Physics C: Solid State Physics}\
  }\textbf {\bibinfo {volume} {10}},\ \bibinfo {pages} {1051--1057} (\bibinfo
  {year} {1977})}\BibitemShut {NoStop}%
\bibitem [{\citenamefont {Nakatsuka}\ \emph {et~al.}(2006)\citenamefont
  {Nakatsuka}, \citenamefont {Ikeda}, \citenamefont {Nakayama},\ and\
  \citenamefont {Mizota}}]{Nakatsuka-CoGa2O4-ref}%
  \BibitemOpen
  \bibfield  {author} {\bibinfo {author} {\bibfnamefont {Akihiko}\ \bibnamefont
  {Nakatsuka}}, \bibinfo {author} {\bibfnamefont {Yuya}\ \bibnamefont {Ikeda}},
  \bibinfo {author} {\bibfnamefont {Noriaki}\ \bibnamefont {Nakayama}}, \ and\
  \bibinfo {author} {\bibfnamefont {Tadato}\ \bibnamefont {Mizota}},\
  }\bibfield  {title} {\enquote {\bibinfo {title} {{Inversion parameter of the
  CoGa${\sb 2}$O${\sb 4}$ spinel determined from single-crystal X-ray data}},}\
  }\href {\doibase 10.1107/S1600536806011044} {\bibfield  {journal} {\bibinfo
  {journal} {Acta Crystallographica Section E}\ }\textbf {\bibinfo {volume}
  {62}},\ \bibinfo {pages} {i109--i111} (\bibinfo {year} {2006})}\BibitemShut
  {NoStop}%
\bibitem [{\citenamefont {Areán}\ and\ \citenamefont
  {Trobajo-Fernandez}(1985)}]{Arean-NiGa2O4}%
  \BibitemOpen
  \bibfield  {author} {\bibinfo {author} {\bibfnamefont {C.~Otero}\
  \bibnamefont {Areán}}\ and\ \bibinfo {author} {\bibfnamefont {M.~C.}\
  \bibnamefont {Trobajo-Fernandez}},\ }\bibfield  {title} {\enquote {\bibinfo
  {title} {Cation distribution in {Mg$_x$Ni$_{1–x}$Ga$_2$O$_4$} oxide
  spinels},}\ }\href {\doibase https://doi.org/10.1002/pssa.2210920213}
  {\bibfield  {journal} {\bibinfo  {journal} {Physica Status Solidi (a)}\
  }\textbf {\bibinfo {volume} {92}},\ \bibinfo {pages} {443--447} (\bibinfo
  {year} {1985})}\BibitemShut {NoStop}%
\bibitem [{\citenamefont {Sasaki}\ \emph {et~al.}(1979)\citenamefont {Sasaki},
  \citenamefont {Fujino},\ and\ \citenamefont
  {Tak\'euchi}}]{rocksaltoxides-structure-nio}%
  \BibitemOpen
  \bibfield  {author} {\bibinfo {author} {\bibfnamefont {Satoshi}\ \bibnamefont
  {Sasaki}}, \bibinfo {author} {\bibfnamefont {Kiyoshi}\ \bibnamefont
  {Fujino}}, \ and\ \bibinfo {author} {\bibfnamefont {Yoshio}\ \bibnamefont
  {Tak\'euchi}},\ }\bibfield  {title} {\enquote {\bibinfo {title} {X-ray
  determination of electron-density distributions in oxides, {MgO, MnO, CoO,
  and NiO}, and atomic scattering factors of their constituent atoms},}\ }\href
  {\doibase 10.2183/pjab.55.43} {\bibfield  {journal} {\bibinfo  {journal}
  {Proceedings of the Japan Academy, Series B}\ }\textbf {\bibinfo {volume}
  {55}},\ \bibinfo {pages} {43--48} (\bibinfo {year} {1979})}\BibitemShut
  {NoStop}%
\bibitem [{Note1()}]{Note1}%
  \BibitemOpen
  \bibinfo {note} {We define $T_C$ by a slope method wherein the approximately
  linear increase in the susceptibility is extrapolated to its
  $x$-intercept.}\BibitemShut {Stop}%
\bibitem [{\citenamefont {Music{\'o}}\ \emph {et~al.}(2019)\citenamefont
  {Music{\'o}}, \citenamefont {Wright}, \citenamefont {Ward}, \citenamefont
  {Grutter}, \citenamefont {Arenholz}, \citenamefont {Gilbert}, \citenamefont
  {Mandrus},\ and\ \citenamefont {Keppens}}]{musico2019tunable}%
  \BibitemOpen
  \bibfield  {author} {\bibinfo {author} {\bibfnamefont {Brianna}\ \bibnamefont
  {Music{\'o}}}, \bibinfo {author} {\bibfnamefont {Quinton}\ \bibnamefont
  {Wright}}, \bibinfo {author} {\bibfnamefont {T~Zac}\ \bibnamefont {Ward}},
  \bibinfo {author} {\bibfnamefont {Alexander}\ \bibnamefont {Grutter}},
  \bibinfo {author} {\bibfnamefont {Elke}\ \bibnamefont {Arenholz}}, \bibinfo
  {author} {\bibfnamefont {Dustin}\ \bibnamefont {Gilbert}}, \bibinfo {author}
  {\bibfnamefont {David}\ \bibnamefont {Mandrus}}, \ and\ \bibinfo {author}
  {\bibfnamefont {Veerle}\ \bibnamefont {Keppens}},\ }\bibfield  {title}
  {\enquote {\bibinfo {title} {Tunable magnetic ordering through cation
  selection in entropic spinel oxides},}\ }\href
  {https://journals.aps.org/prmaterials/abstract/10.1103/PhysRevMaterials.3.104416}
  {\bibfield  {journal} {\bibinfo  {journal} {Physical Review Materials}\
  }\textbf {\bibinfo {volume} {3}},\ \bibinfo {pages} {104416} (\bibinfo {year}
  {2019})}\BibitemShut {NoStop}%
\bibitem [{\citenamefont {Cieslak}\ \emph
  {et~al.}(2021{\natexlab{b}})\citenamefont {Cieslak}, \citenamefont
  {Reissner}, \citenamefont {Berent}, \citenamefont {Dabrowa}, \citenamefont
  {Stygar}, \citenamefont {Mozdzierz},\ and\ \citenamefont
  {Zajusz}}]{cieslak2021magnetic}%
  \BibitemOpen
  \bibfield  {author} {\bibinfo {author} {\bibfnamefont {J}~\bibnamefont
  {Cieslak}}, \bibinfo {author} {\bibfnamefont {M}~\bibnamefont {Reissner}},
  \bibinfo {author} {\bibfnamefont {K}~\bibnamefont {Berent}}, \bibinfo
  {author} {\bibfnamefont {J}~\bibnamefont {Dabrowa}}, \bibinfo {author}
  {\bibfnamefont {M}~\bibnamefont {Stygar}}, \bibinfo {author} {\bibfnamefont
  {M}~\bibnamefont {Mozdzierz}}, \ and\ \bibinfo {author} {\bibfnamefont
  {M}~\bibnamefont {Zajusz}},\ }\bibfield  {title} {\enquote {\bibinfo {title}
  {Magnetic properties and ionic distribution in high entropy spinels studied
  by {M{\"o}ssbauer} and ab initio methods},}\ }\href
  {https://doi.org/10.1016/j.actamat.2020.116600} {\bibfield  {journal}
  {\bibinfo  {journal} {Acta Materialia}\ }\textbf {\bibinfo {volume} {206}},\
  \bibinfo {pages} {116600} (\bibinfo {year} {2021}{\natexlab{b}})}\BibitemShut
  {NoStop}%
\bibitem [{\citenamefont {Sucksmith}\ and\ \citenamefont
  {Thompson}(1954)}]{sucksmith1954magnetic}%
  \BibitemOpen
  \bibfield  {author} {\bibinfo {author} {\bibfnamefont {Willie}\ \bibnamefont
  {Sucksmith}}\ and\ \bibinfo {author} {\bibfnamefont {Jo~E}\ \bibnamefont
  {Thompson}},\ }\bibfield  {title} {\enquote {\bibinfo {title} {The magnetic
  anisotropy of cobalt},}\ }\href {https://doi.org/10.1098/rspa.1954.0209}
  {\bibfield  {journal} {\bibinfo  {journal} {Proceedings of the Royal Society
  of London. Series A. Mathematical and Physical Sciences}\ }\textbf {\bibinfo
  {volume} {225}},\ \bibinfo {pages} {362--375} (\bibinfo {year}
  {1954})}\BibitemShut {NoStop}%
\bibitem [{\citenamefont {Lines}(1963)}]{lines1963magnetic}%
  \BibitemOpen
  \bibfield  {author} {\bibinfo {author} {\bibfnamefont {ME}~\bibnamefont
  {Lines}},\ }\bibfield  {title} {\enquote {\bibinfo {title} {Magnetic
  properties of {CoCl$_2$} and {NiCl$_2$}},}\ }\href
  {https://journals.aps.org/pr/abstract/10.1103/PhysRev.131.546} {\bibfield
  {journal} {\bibinfo  {journal} {Physical Review}\ }\textbf {\bibinfo {volume}
  {131}},\ \bibinfo {pages} {546} (\bibinfo {year} {1963})}\BibitemShut
  {NoStop}%
\bibitem [{\citenamefont {Yafet}\ and\ \citenamefont
  {Kittel}(1952)}]{yafet1952antiferromagnetic}%
  \BibitemOpen
  \bibfield  {author} {\bibinfo {author} {\bibfnamefont {Y}~\bibnamefont
  {Yafet}}\ and\ \bibinfo {author} {\bibfnamefont {Ch}~\bibnamefont {Kittel}},\
  }\bibfield  {title} {\enquote {\bibinfo {title} {Antiferromagnetic
  arrangements in ferrites},}\ }\href
  {https://journals.aps.org/pr/abstract/10.1103/PhysRev.87.290} {\bibfield
  {journal} {\bibinfo  {journal} {Physical Review}\ }\textbf {\bibinfo {volume}
  {87}},\ \bibinfo {pages} {290} (\bibinfo {year} {1952})}\BibitemShut
  {NoStop}%
\bibitem [{\citenamefont {Murthy}\ \emph {et~al.}(1969)\citenamefont {Murthy},
  \citenamefont {Natera}, \citenamefont {Youssef}, \citenamefont {Begum},\ and\
  \citenamefont {Srivastava}}]{murthy1969yafet}%
  \BibitemOpen
  \bibfield  {author} {\bibinfo {author} {\bibfnamefont {NS~Satya}\
  \bibnamefont {Murthy}}, \bibinfo {author} {\bibfnamefont {MG}~\bibnamefont
  {Natera}}, \bibinfo {author} {\bibfnamefont {SI}~\bibnamefont {Youssef}},
  \bibinfo {author} {\bibfnamefont {RJ}~\bibnamefont {Begum}}, \ and\ \bibinfo
  {author} {\bibfnamefont {CM}~\bibnamefont {Srivastava}},\ }\bibfield  {title}
  {\enquote {\bibinfo {title} {{Yafet-Kittel} angles in zinc-nickel
  ferrites},}\ }\href
  {https://journals.aps.org/pr/abstract/10.1103/PhysRev.181.969} {\bibfield
  {journal} {\bibinfo  {journal} {Physical review}\ }\textbf {\bibinfo {volume}
  {181}},\ \bibinfo {pages} {969} (\bibinfo {year} {1969})}\BibitemShut
  {NoStop}%
\bibitem [{\citenamefont {Willard}\ \emph {et~al.}(1999)\citenamefont
  {Willard}, \citenamefont {Nakamura}, \citenamefont {Laughlin},\ and\
  \citenamefont {McHenry}}]{willard1999magnetic}%
  \BibitemOpen
  \bibfield  {author} {\bibinfo {author} {\bibfnamefont {Matthew~A}\
  \bibnamefont {Willard}}, \bibinfo {author} {\bibfnamefont {Yuichiro}\
  \bibnamefont {Nakamura}}, \bibinfo {author} {\bibfnamefont {David~E}\
  \bibnamefont {Laughlin}}, \ and\ \bibinfo {author} {\bibfnamefont
  {Michael~E}\ \bibnamefont {McHenry}},\ }\bibfield  {title} {\enquote
  {\bibinfo {title} {Magnetic properties of ordered and disordered spinel-phase
  ferrimagnets},}\ }\href {https://doi.org/10.1111/j.1151-2916.1999.tb02249.x}
  {\bibfield  {journal} {\bibinfo  {journal} {Journal of the American Ceramic
  Society}\ }\textbf {\bibinfo {volume} {82}},\ \bibinfo {pages} {3342--3346}
  (\bibinfo {year} {1999})}\BibitemShut {NoStop}%
\bibitem [{\citenamefont {Pattrick}\ \emph {et~al.}(2002)\citenamefont
  {Pattrick}, \citenamefont {Van Der~Laan}, \citenamefont {Henderson},
  \citenamefont {Kuiper}, \citenamefont {Dudzik},\ and\ \citenamefont
  {Vaughan}}]{Pattrick-VanDerLaan-cation-occupancy-spinels}%
  \BibitemOpen
  \bibfield  {author} {\bibinfo {author} {\bibfnamefont {Richard~A.D.}\
  \bibnamefont {Pattrick}}, \bibinfo {author} {\bibfnamefont {Gerrit}\
  \bibnamefont {Van Der~Laan}}, \bibinfo {author} {\bibfnamefont
  {C.~Michael~B.}\ \bibnamefont {Henderson}}, \bibinfo {author} {\bibfnamefont
  {Pieter}\ \bibnamefont {Kuiper}}, \bibinfo {author} {\bibfnamefont {Esther}\
  \bibnamefont {Dudzik}}, \ and\ \bibinfo {author} {\bibfnamefont {David~J.}\
  \bibnamefont {Vaughan}},\ }\bibfield  {title} {\enquote {\bibinfo {title}
  {Cation site occupancy in spinel ferrites studied by x-ray magnetic circular
  dichroism: developing a method for mineralogists},}\ }\href {\doibase
  10.1127/0935-1221/2002/0014-1095} {\bibfield  {journal} {\bibinfo  {journal}
  {European Journal of Mineralogy}\ }\textbf {\bibinfo {volume} {14}},\
  \bibinfo {pages} {1095--1102} (\bibinfo {year} {2002})}\BibitemShut {NoStop}%
\bibitem [{\citenamefont {van Schooneveld}\ \emph {et~al.}(2012)\citenamefont
  {van Schooneveld}, \citenamefont {Kurian}, \citenamefont {Juhin},
  \citenamefont {Zhou}, \citenamefont {Schlappa}, \citenamefont {Strocov},
  \citenamefont {Schmitt},\ and\ \citenamefont
  {de~Groot}}]{Multiplet-Co-DeGroot}%
  \BibitemOpen
  \bibfield  {author} {\bibinfo {author} {\bibfnamefont {Matti~M.}\
  \bibnamefont {van Schooneveld}}, \bibinfo {author} {\bibfnamefont {Reshmi}\
  \bibnamefont {Kurian}}, \bibinfo {author} {\bibfnamefont {Amélie}\
  \bibnamefont {Juhin}}, \bibinfo {author} {\bibfnamefont {Kejin}\ \bibnamefont
  {Zhou}}, \bibinfo {author} {\bibfnamefont {Justine}\ \bibnamefont
  {Schlappa}}, \bibinfo {author} {\bibfnamefont {Vladimir~N.}\ \bibnamefont
  {Strocov}}, \bibinfo {author} {\bibfnamefont {Thorsten}\ \bibnamefont
  {Schmitt}}, \ and\ \bibinfo {author} {\bibfnamefont {Frank M.~F.}\
  \bibnamefont {de~Groot}},\ }\bibfield  {title} {\enquote {\bibinfo {title}
  {Electronic structure of {CoO} nanocrystals and a single crystal probed by
  resonant x-ray emission spectroscopy},}\ }\href {\doibase 10.1021/jp302847h}
  {\bibfield  {journal} {\bibinfo  {journal} {The Journal of Physical Chemistry
  C}\ }\textbf {\bibinfo {volume} {116}},\ \bibinfo {pages} {15218--15230}
  (\bibinfo {year} {2012})}\BibitemShut {NoStop}%
\bibitem [{\citenamefont {Hibberd}\ \emph {et~al.}(2015)\citenamefont
  {Hibberd}, \citenamefont {Doan}, \citenamefont {Glass}, \citenamefont
  {de~Groot}, \citenamefont {Hill},\ and\ \citenamefont {Cuk}}]{Co-2-JT-ref}%
  \BibitemOpen
  \bibfield  {author} {\bibinfo {author} {\bibfnamefont {Amber~M.}\
  \bibnamefont {Hibberd}}, \bibinfo {author} {\bibfnamefont {Hoang~Q.}\
  \bibnamefont {Doan}}, \bibinfo {author} {\bibfnamefont {Elliot~N.}\
  \bibnamefont {Glass}}, \bibinfo {author} {\bibfnamefont {Frank M.~F.}\
  \bibnamefont {de~Groot}}, \bibinfo {author} {\bibfnamefont {Craig~L.}\
  \bibnamefont {Hill}}, \ and\ \bibinfo {author} {\bibfnamefont {Tanja}\
  \bibnamefont {Cuk}},\ }\bibfield  {title} {\enquote {\bibinfo {title} {Co
  polyoxometalates and a {Co$_3$O$_4$} thin film investigated by {L}-edge x-ray
  absorption spectroscopy},}\ }\href {\doibase 10.1021/jp5124037} {\bibfield
  {journal} {\bibinfo  {journal} {The Journal of Physical Chemistry C}\
  }\textbf {\bibinfo {volume} {119}},\ \bibinfo {pages} {4173--4179} (\bibinfo
  {year} {2015})}\BibitemShut {NoStop}%
\end{thebibliography}%

\end{document}